\RequirePackage{rotating}
\documentclass[fleqn,usenatbib]{mnras}

\usepackage{mathptmx}
\usepackage{array, longtable}
\usepackage{caption}
\usepackage{rotating}
\usepackage{hyperref}
\usepackage{adjustbox}
\usepackage{placeins}
\usepackage{xcolor}

\usepackage[T1]{fontenc}
\usepackage{pifont}

\DeclareRobustCommand{\VAN}[3]{#2}
\let\VANthebibliography\thebibliography
\def\thebibliography{\DeclareRobustCommand{\VAN}[3]{##3}\VANthebibliography}

\usepackage{graphicx}	
\usepackage{amsmath}	
\usepackage{amssymb}	
\usepackage{xcolor}

\title[X-Shooter Survey of Young IMSs]{X-Shooter Survey of Young Intermediate Mass Stars - I. Stellar Characterization and Disc Evolution}

\author[D. P. Iglesias et al.]{Daniela P. Iglesias$^{1}$,\thanks{E-mail: D.P.Iglesias@leeds.ac.uk (DPI)}
Olja Pani\'c$^{1}$,
Mario van den Ancker,$^{2}$
Monika G. Petr-Gotzens$^{2}$,
Lionel Siess$^{3}$,
\newauthor 
Miguel Vioque$^{4,5}$,
Ilaria Pascucci$^{6,7}$,
Ren\'e Oudmaijer$^{1}$,
James Miley$^{4,8}$
\\
\\
$^{1}$School of Physics and Astronomy, Sir William Henry Bragg Building, University of Leeds, Leeds LS2 9JT, UK\\
$^{2}$European Southern Observatory, Karl-Schwarzschild-Str. 2, D–85748 Garching bei München, Germany \\
$^{3}$Institut d’Astronomie et d’Astrophysique, Universit\'e Libre de Bruxelles (ULB), CP 226, 1050 Brussels, Belgium \\
$^{4}$Joint ALMA Observatory, Alonso de C\' ordova 3107, Vitacura, Santiago 763-0355, Chile \\
$^{5}$National Radio Astronomy Observatory, 520 Edgemont Road, Charlottesville, VA 22903, USA \\
$^{6}$ Lunar and Planetary Laboratory, The University of Arizona, Tucson, AZ 85721, USA \\
$^{7}$ Earths in Other Solar Systems Team, NASA Nexus for Exoplanet System Science, USA \\
$^{8}$National Astronomical Observatory of Japan (NAOJ), Los Abedules 3085, Office 701, Vitacura, Santiago, Chile
}

\date{Accepted XXX. Received YYY; in original form ZZZ}
\pubyear{2022}
\begin{document}
\label{firstpage}
\pagerange{\pageref{firstpage}--\pageref{lastpage}}
\maketitle

\begin{abstract}
Intermediate mass stars (IMSs) represent the link between low-mass and high-mass stars and cover a key mass range for giant planet formation. In this paper, we present a spectroscopic survey of 241 young IMS candidates with IR-excess, the most complete unbiased sample to date within 300 pc. We combined VLT/X-Shooter spectra with BVR photometric observations and Gaia DR3 distances to estimate fundamental stellar parameters such as T$_{\rm eff}$, mass, radius, age, and luminosity. We further selected those stars within the intermediate mass range $1.5 \leq M_{\star}/M_{\odot} \leq 3.5$ and discarded old contaminants. We used 2MASS and WISE photometry to study the IR-excesses of the sample, finding 92 previously unidentified stars with IR-excess. We classified this sample into `protoplanetary', `hybrid candidates' and `debris' discs based on their observed fractional excess at 12 $\mu$m, finding a new population of 17 hybrid disc candidates. We studied inner disc dispersal timescales for $\lambda < 10 \mu$m and found very different trends for IMSs and low mass stars (LMSs). IMSs show excesses dropping fast during the first 6 Myrs independently of the wavelength, while LMSs show consistently lower fractions of excess at the shortest wavelengths and increasingly higher fractions for longer wavelengths, with slower dispersal rates. 
In conclusion, this study demonstrates empirically that IMSs dissipate their inner discs very differently than LMSs, providing a possible explanation for the lack of short period planets around IMSs.

\end{abstract}




\section{Introduction}

The discovery of thousands of planets around different types of stars has allowed astronomers to study a variety of planetary systems architectures and obtain demographics on exoplanets populations (e.g. \citealt{Mulders2018}, \citealt{Yang2020}, \citealt{Gaudi2022}). These statistics give hints on how planet formation depends on the stellar characteristics of the host star, such as mass and multiplicity. In particular, there is evidence that the occurrence rate of giant planets peak around intermediate mass stars  \citep{Lovis_Mayor2007,Reffert2015}. Interestingly, giant planet frequency appears to drop dramatically for higher mass stars ($>$3.5M$_{\odot}$), and this might be due to their faster disc dispersal \citep{Pinilla2022}. Intermediate mass stars (IMSs: stars with masses within the range $1.5 \leq M_{\star}/M_{\odot} \leq 3.5$) seem to offer a "sweet spot" for giant planet formation given their particular disc sizes, disc masses, and disc dispersal rates. The distribution of planets around IMSs seems to concentrate at 1--2 au from the star, however there is a lack of low-mass (M$_{\rm planet} <$ 0.7M$_{\rm Jup}$) short period planets with semimajor axes $<$0.6 au \citep{Johnson2007b, Bowler2010, Moe2021}. Although planets are more difficult to detect around main-sequence IMSs than in low mass stars (LMSs) due to their intrinsic stellar properties (higher luminosity, pulsations, faster rotation leading to broadened spectral lines), radial velocity surveys have been carried around more evolved IMSs which have slower rotational speed and thus narrower lines, and yet this so called ``planet desert'' of short period low-mass planets remains \citep{Johnson2010, Reffert2015, Medina2018}. In addition, since short period planets (P$<$20d) should be easier to detect via transit than long period ones as they are more likely to transit the star and transit more often, then these planets must be intrinsically rare around IMSs. Theoretical studies demonstrate that giant planet formation in IMSs is more efficient than in low mass stars. Planetary embryos around IMSs rapidly accumulate enough mass for their cores to be able to accrete gas before the disc gets dispersed \citep{Kretke2009}. The favoured location for these planetary embryos to grow is the inner edge of the "dead zone"; an inactive region trapped between the inner region of the disc, where the gas is thermally ionized, and turbulent layers further out in the disc, where the surfaces become ionized by other phenomena such as X-rays or cosmic rays \citep{Kretke2009}.  

The properties of circumstellar discs, as the building blocks of planets, are of course determinant in the architecture of planetary systems. However, their structure is also dependant on their stellar host. Disc mass is one of the most relevant properties as it basically determines the amount of material available for building planets and, as a consequence, has an impact on the size and number of planets in a system. Studies show that the disc mass--stellar mass dependence is steeper than linear and that dust masses in IMSs are significantly higher than in low mass stars (\citealt{Pascucci2016}, \citealt{Stapper2022}). This is thought to be related to the frequency of giant planets around IMSs. However, not all the disc material will be destined to build planets; part of it will accrete onto the star and part of it will be lost from the system via winds and outflows. As giant planets can only form and migrate in presence of dense gas before the disc is depleted, it is crucial to understand the when and the how of disc dispersal  (e.g., \citealt{Nelson2004}, \citealt{Paardekooper2009}).

Most disc dispersal studies have focused on LMSs given their high frequency among stellar populations (e.g. \citealt{Haisch2001}, \citealt{Hernandez2010}, \citealt{Ercolano2011}). Other works, such as \cite{Ribas2015} compared disc evolution in LMSs against IMSs together with very massive stars up to 8 M$_{\odot}$, which are known to evolve very differently (e.g. \citealt{Woosley2002}). Recent surveys of IMSs have focused in particular on Herbig Ae/Be type stars (e.g. \citealt{Wichittanakom2020}, \citealt{Vioque2020, Vioque2022}). These are also pre-main-sequence IMSs, however, they are undergoing gas accretion and thus are not ideal targets for the study of inner disc clearing.



Several key differences are found between LMSs and IMSs after pre-main sequence. For example, IMSs are the most frequent hosts of giant planets, both at short and large orbital separations (\citealt{Reffert2015}, \citealt{Wagner2022}). IMSs also host the brightest debris discs, which in a number of cases are coincident with directly imaged planets (\citealt{Chauvin2018}, \citealt{Lagrange2019}, \citealt{Marois2008}). Perhaps the most pertinent difference to the contents of this paper is the fact that gas in debris discs is found almost exclusively around IMSs. Gas detection rate in debris discs around main sequence A-type IMSs stars is nearly 70\%, while for lower mass F-G-K stars this falls to only 7\% \citep{Moor2017}. This strongly implies that the physical mechanism responsible for the presence of gas in debris discs must operate preferentially around IMSs, and might be directly related to the gas disc lifetime in IMSs \citep{Nakatani2021}. 

In the present paper we aim to study the inner disc evolution of IMSs with the purpose of better understanding the particular process of planet formation around IMSs. For this, we have collected the largest, unbiased sample of young IMSs surrounded by circumstellar discs. The best way to measure the clearing of the inner disc regions (up to a few au) is by studying the near-IR excess caused by dust.  In this study, we accurately measure the basic stellar parameters of the sample, study their near-to-mid infra-red excesses, and compare disc evolution in IMSs to studies of low mass stars.

This paper is structured as follows: in Section 2 we describe our initial sample of study, the observations used in this work and their reduction. Section 3 explains the methodology used in this work to estimate all the basic stellar parameters for the sample, such as T$_{\rm eff}$, masses and ages, and the calculation of infra-red excesses at different wavelengths. In this section we also further refine the initial sample and perform an assessment of the ages to discard evolved contaminants. In Section 4 we perform an analysis and discussion of the results obtained in Section 3. We conduct a disc classification based on the IR-excess levels at 12 $\mu$m and make several comparison to previous studies of low mass stars and others studies on disc evolution. In Section 5 we present the conclusions, and in the Appendix we present additional results as byproducts of the survey such as accretion diagnostics for the Herbig Ae/Be stars found in the sample and some comments on individual objects.

\section{SAMPLE SELECTION, OBSERVATIONS AND DATA REDUCTION}

\subsection{Sample Selection}
\label{sec:sample} 

We used the Tycho-Gaia Astrometric Solution (TGAS) \citep{Gaia2016} and Hipparcos catalogues \citep{VanLeeuwen2007} to select new intermediate-mass pre-main sequence stars candidates. The candidates are spatially distributed all over the southern sky as observable from Paranal observatory in Chile and not concentrated in any particular star forming region. Selecting only the stars which have TGAS B$_{T}$ and V$_{T}$ photometry, we used the effective temperature T$_{\rm eff}$ values given in the Tycho-2 Spectral Type Catalogue \citep{Wright2003} and the parallaxes from TGAS (or Hipparcos when TGAS were not available), and estimated stellar luminosities. Luminosities were obtained by fitting black body profiles of the appropriate T$_{\rm eff}$ to the dereddened photometry. These allowed us to place the sources in an HR diagram, where we selected only those stars consistent with $1.5 \leq M_{\star}/M_{\odot} \leq 3.5$. 
This selection was then crossmatched with the WISE catalogue to look for mid-IR excesses. Only stars with measured excess levels W1-W4 $\geq 1$ were kept. This limit corresponds to a debris disc level of excess at $22\mu$m \citep{Wyatt2015}. 
The sample was volume limited by setting the furthest distance at 300 pc. All the catalogues we used are limited in magnitude, being the brightest limit V$_{\rm mag}<\sim11$, but because our stars are inherently bright, these biases do not affect our sample within the volume selection. We estimate that our sample is 35-55\% complete within 300pc for sources $<$5 Myr and 1.5 to 3.5 $M_{\odot}$.

 The candidates selection consists of 241 objects, of which 27 had previous X-Shooter archival data. The candidates and their individual coordinates are listed in Table \ref{tab:observations} and relevant information taken from the literature such as spectral types, photometry and distances is presented in Table \ref{tab:liter}.

\begin{table*}
	\centering
	\caption{Full sample of pre-selected young intermediate mass stars candidates. Columns 1 to 9 list identifier used for this work (either HIP or TYC), Simbad main identifier, coordinates, number of observations, observing date (or first and last observing date for the case of more than one epoch), and SNR achieved for the UVB, VIS, and NIR arms of X-Shooter (maximum value, if more than one epoch).}
	\label{tab:observations}
	\begin{tabular*}{\textwidth}{lcccccccc} 
		\hline
		Name & Simbad ID & RA & DEC & N obs. & Obs. date & SNR(UVB) & SNR(VIS) & SNR(NIR) \\
		     & & (J2000) & (J2000) & & (yyyy-mm-dd) & & & \\
		\hline
HIP\,4088 & HD\,5208 & 00:52:28.35 & -69:30:10.4 & 1 & 2018-07-16 & 648 & 730 & 459 \\ 
TYC\,628-841-1 & BD+13\,277 & 01:47:57.86 & +13:53:09.4 & 1 & 2018-06-14 & 91 & 104 & 32 \\ 
TYC\,6433-1491-1 & CD-24\,991 & 02:16:23.15 & -23:49:36.0 & 1 & 2018-07-25 & 238 & 244 & 108 \\ 
TYC\,1765-1373-1 & BD+22\,333B & 02:21:36.14 & +23:37:54.7 & 1 & 2018-10-12 & 206 & 208 & 83 \\ 
HIP\,12055 & HD\,16152 & 02:35:24.47 & -09:21:02.7 & 1 & 2018-07-28 & 778 & 590 & 280 \\ 
HIP\,13910 & HD\,18572 & 02:59:08.41 & -04:47:00.1 & 1 & 2018-07-28 & 716 & 508 & 268 \\ 
TYC\,56-284-1 & HD\,20246 & 03:15:25.07 & +01:41:46.7 & 1 & 2018-10-22 & 399 & 380 & 262 \\ 
TYC\,1235-186-1 & BD+14\,592 & 03:40:14.42 & +15:11:26.3 & 1 & 2018-10-22 & 395 & 560 & 496 \\ 
HIP\,17527 & *18\,Tau & 03:45:09.74 & +24:50:21.3 & 7 & 2015-10-27 -- 2018-10-19 & 1011 & 713 & 629 \\ 
HIP\,17543 & V*\,CT\,Hyi & 03:45:23.74 & -71:39:29.3 & 2 & 2009-10-15 -- 2018-10-22 & 710 & 492 & 151 \\  
... & ... & ... & ... & ... & ... & ... & ... & ... \\
		\hline
	\end{tabular*}
		\begin{description}
      \item \textbf{Notes:} Only a portion of this table is presented here to show its content and description. Given its size (241 objects), the full table is only available at the CDS. Coordinates and Simbad ID were taken from Simbad database \citep{Wenger2000}. 
	\end{description}
\end{table*}
\begin{table*}
	\centering
	\caption{Information taken from the literature for the full sample of pre-selected IMSs. Columns 1 and 2 list identifier and spectral types. Columns 3 to 10 present the photometry used in this study; B$_{\rm T}$V$_{\rm T}$R photometry, $2MASS$ K$_{s}$ and the four WISE bands. Column 11 shows distances derived from Gaia DR3 parallaxes, except otherwise indicated.}
	\label{tab:liter}
	\begin{tabular*}{\textwidth}{lcccccccccc} 
		\hline
			Name  & Spectral & B$_{\rm T}$ & V$_{\rm T}$ & R & K$_{s}$ & W1 & W2 & W3  & W4 & Dist.    \\
			   &   type   & (mag) & (mag) & (mag) & (mag) & ( mag) & (mag) & (mag) & (mag) & (pc)      \\
		\hline
HIP\,4088 & G0/2V & 7.89 & 7.29 & 6.88 & 5.81$\pm$0.03 & 6.78$\pm$0.12 & 7.77 (u.l.) & 5.81$\pm$0.02 & 5.75$\pm$0.05 & 60.22$\pm^{0.10}_{0.10}$ \\ 
TYC\,628-841-1 & G5 & 11.57 & 10.81 & 10.32 & 9.53$\pm$0.02 & 9.48$\pm$0.02 & 9.52$\pm$0.02 & 9.31$\pm$0.04 & 8.32$\pm$0.26 & 237.81$\pm^{1.13}_{0.75}$ \\ 
TYC\,6433-1491-1 & G5 & 11.20 & 10.59 & 10.18 & 9.40$\pm$0.02 & 9.38$\pm$0.02 & 9.41$\pm$0.02 & 9.27$\pm$0.04 & 8.26$\pm$0.22 & 199.04$\pm^{0.52}_{0.58}$ \\ 
TYC\,1765-1373-1 & K0 & 10.89 & 10.22 & 9.77 & 9.10$\pm$0.02 & 8.81$\pm$0.03 & 8.77$\pm$0.02 & 8.31$\pm$0.03 & 7.18$\pm$0.12 & 257.41$\pm^{1.30}_{1.39}$ \\ 
HIP\,12055 & A0V & 7.13 & 7.10 & 7.09 & 7.06$\pm$0.02 & 7.02$\pm$0.05 & 7.04$\pm$0.02 & 6.72$\pm$0.01 & 5.59$\pm$0.04 & 131.35$\pm^{0.49}_{0.54}$ \\ 
HIP\,13910 & B9.5V & 8.04 & 8.02 & 8.01 & 8.00$\pm$0.02 & 7.92$\pm$0.02 & 7.95$\pm$0.02 & 7.73$\pm$0.02 & 6.33$\pm$0.06 & 217.28$\pm^{2.11}_{1.84}$ \\ 
TYC\,56-284-1 & F7V & 9.33 & 8.89 & 8.60 & 7.69$\pm$0.02 & 7.50$\pm$0.03 & 7.53$\pm$0.02 & 7.18$\pm$0.02 & 6.50$\pm$0.07 & 161.08$\pm^{0.77}_{0.79}$ \\ 
TYC\,1235-186-1 & K0 & 10.18 & 9.15 & 8.51 & 6.56$\pm$0.02 & 7.45$\pm$0.40 & 6.46$\pm$0.02 & 6.49$\pm$0.02 & 6.43$\pm$0.07 & 152.68$\pm^{0.46}_{0.54}$ \\ 
*HIP\,17527 & B7V & 5.58 & 5.64 & 5.68 & 5.81$\pm$0.02 & 5.83$\pm$0.12 & 5.72$\pm$0.04 & 5.55$\pm$0.01 & 3.84$\pm$0.03 & 137.86$\pm^{1.05}_{0.81}$ \\ 
*HIP\,17543 & ApSi & 6.14 & 6.25 & 6.31 & 6.58$\pm$0.03 & 6.59$\pm$0.08 & 6.56$\pm$0.02 & 6.15$\pm$0.01 & 4.24$\pm$0.02 & 153.28$\pm^{0.69}_{0.79}$ \\ 
... & ... & ... & ... & ... & ... & ... & ... & ... & ... & ... \\
		\hline
	\end{tabular*}
		\begin{description}
      \item \textbf{Notes:} Only a portion of this table is presented here to show its content, description and references. Given its size, the full table is only available at the CDS. Spectral types and B$_{\rm T}$V$_{\rm T}$ photometry were taken from the Tycho-2 Spectral Type Catalogue \citep{Wright2003}. R photometry was obtained from the USNO-B Catalog \citep{Monet2003}. Distances were taken from \cite{Bailer-Jones2021}, except for a few cases not available in DR3: HIP\,57027 taken from \cite{GaiaDR2_2018}, and HIP\,59896, HIP\,69761, HIP\,74911, and HIP\,91893, taken from \cite{VanLeeuwen2007}. Objects with previous studies in the literature are marked with a `*'.
	\end{description}
\end{table*}

\subsection{Observations}
\label{sec:obs} 

We carried out observations with the X-Shooter echelle spectrograph \citep{Vernet2011} for 214 objects in the sample during ESO periods 101 and 102\footnote{Under ESO programmes 0101.C-0902(A) and 0102.C-0882(A), respectively.} (i.e. between April 1018 and March 2019). X-Shooter is mounted at the Very Large Telescope (VLT) and covers a total wavelength range of 3000–23000\AA\, split into three arms: the UV-blue (UVB) arm covering 3000–5600\AA; the visible (VIS) arm covering 5500–10200\AA; and the near-IR (NIR) arm covering 10200–24800\AA. Observations were performed using the 1.0'', 0.9'', and 0.6'' slit widths for the UVB, VIS and NIR arms, respectively, resulting in the corresponding spectral resolutions of 5400, 8900 and 8100. Exposure times were estimated using the X-Shooter Exposure Time Calculator\footnote{\url{https://www.eso.org/observing/etc/bin/gen/form?INS.NAME=X-SHOOTER+INS.MODE=spectro}} with the intent of achieving a signal-to-noise ratio SNR$>$100 for the three X-Shooter arms respectively. Since our targets are bright (V$_{mag}$: 4.7 -- 11.4) the required SNR was achieved in 120 seconds or less. The total exposure times were split in four exposures (combined into a single spectrum during the reduction process), and for the NIR one nodding cycle was used to ensure good sky subtraction.

For the remaining 27 objects in our sample, we collected the already available X-Shooter data from the ESO archive to complete our database of X-Shooter observations for the full sample. 
The number of observations obtained for each object, observing dates, and SNR are summarized in Table \ref{tab:observations}.

\subsection{Data Reduction}
\label{sec:reduc} 

We reduced the obtained X-Shooter data using the version 3.5.0 of the X-Shooter EsoReflex pipeline \citep{Modigliani2010,Freudling2013}. 
In a few cases where data was affected by saturation, the saturated exposures were removed from the pipeline input (typically four exposures were combined) and only the best quality data were co-added for the final product. For the cases where too many bad pixels were present within a region of the 2D data and a gap was produced in the extracted spectrum, we increased the interpolation range in the pipeline setup to allow it to construct the spectrum profile over a wider wavelength range interpolating between bad pixels.
Telluric corrections were performed for all the spectra using Molecfit (\citealt{Smette2015}, \citealt{Kausch2015}). This is particularly important for the VIS and NIR ranges as they get significantly affected by contamination from H$_{2}$O, O$_{2}$, O$_{3}$ and CO$_{2}$ (among other molecules) in the Earth's atmosphere. Barycentric radial velocity corrections were also applied a posteriori to the spectra. This way we are able to measure the stellar radial velocities with respect to the barycentric rest frame (see Section \ref{sec:RV}). 

\clearpage
\begin{sidewaystable}
	\centering
	\small
	\caption{Stellar parameters estimations for all the candidates. Columns 1 to 14 list identifier, radial velocity, $v$sin$i$, T$_{\rm eff}$, $\log g$, extinction $A_{V}$, $D/R_{\star}$ distance/radius ratio, stellar radius, luminosity, pre-ZAMS estimation of mass and age, post-ZAMS estimation of mass and age, and a flag for the most likely values; `pre' or `post' ZAMS. Uncertainties for $v$sin$i$ are half the step size used in the grid of models; i.e. 5 km s$^{-1}$.}
	\label{tab:params}
    \begin{tabular*}{\textwidth}{lc@{\extracolsep{\fill}}c@{\extracolsep{\fill}}ccccccccccccccc} 
		\hline
		Name & RV  & $v$sin$i$ & T$_{\rm eff}$ & $\log g$  & $A_{V}$ & D/R$_{\star}$ & R$_{\star}$ &  L$_{\star}$ & M$_{\star,pre-MS}$ & Age$_{pre-MS}$ & M$_{\star,post-MS}$ & Age$_{post-MS}$ & flag \\
		 & (km s$^{-1}$) & (km s$^{-1}$) & (K) & (dex) & (mag)  & (pc/$R_{\odot}$) & ($R_{\odot}$) & ($L_{\odot}$) & ($M_{\odot}$)  & (Myrs) & ($M_{\odot}$)  & (Myrs) &  \\
		\hline
HIP\,4088 & 29$\pm$3 & 40 & 6500$\pm$300 & 4.7$\pm$0.6 & 0.37$\pm^{0.24}_{0.25}$ & 36.21$\pm^{1.16}_{1.07}$ & 1.66$\pm^{0.05}_{0.05}$ & 4.3$\pm^{0.8}_{0.8}$ & 1.36$\pm^{0.16}_{0.05}$ & 13$\pm^{3}_{4}$ & 1.33$\pm^{0.08}_{0.09}$ & 2900$\pm^{1700}_{600}$  & pre \\ 
TYC\,628-841-1 & -22$\pm$3 & 40 & 6500$\pm$300 & 4.6$\pm$0.6 & 0.80$\pm^{0.24}_{0.25}$ & 150.40$\pm^{4.82}_{4.46}$ & 1.58$\pm^{0.05}_{0.05}$ & 4.0$\pm^{0.7}_{0.7}$ & 1.37$\pm^{0.07}_{0.06}$ & 15$\pm^{2}_{4}$ & 1.31$\pm^{0.10}_{0.13}$ & 3200$\pm^{2700}_{900}$  & pre \\ 
TYC\,6433-1491-1 & -15$\pm$3 & 40 & 6500$\pm$200 & 4.8$\pm$0.5 & 0.39$\pm^{0.24}_{0.24}$ & 163.72$\pm^{4.52}_{4.92}$ & 1.22$\pm^{0.04}_{0.03}$ & 2.4$\pm^{0.3}_{0.3}$ & 1.27$\pm^{0.02}_{0.05}$ & 20$\pm^{9}_{1}$ & 1.23$\pm^{0.06}_{0.08}$ & 3200$\pm^{2300}_{600}$  & pre \\ 
TYC\,1765-1373-1 & 1$\pm$3 & 50 & 7100$\pm$400 & 4.5$\pm$0.7 & 1.19$\pm^{0.17}_{0.42}$ & 120.35$\pm^{5.32}_{0.81}$ & 2.14$\pm^{0.02}_{0.10}$ & 11$\pm^{2}_{2}$ & 1.66$\pm^{0.12}_{0.10}$ & 9$\pm^{2}_{2}$ & 1.63$\pm^{0.16}_{0.07}$ & 1700$\pm^{600}_{600}$  & pre \\ 
HIP\,12055 & 17$\pm$3 & 170 & 10700$\pm$600 & 4.4$\pm$0.6 & 0.50$\pm^{0.04}_{0.04}$ & 71.85$\pm^{1.99}_{2.20}$ & 1.83$\pm^{0.06}_{0.05}$ & 40$\pm^{9}_{9}$ & 2.37$\pm^{0.20}_{0.14}$ & 5$\pm^{1}_{1}$ & 2.39$\pm^{0.18}_{0.13}$ & 610$\pm^{25}_{80}$  & pre \\ 
HIP\,13910 & 12$\pm$2 & 220 & 11300$\pm$700 & 4.3$\pm$0.5 & 0.51$\pm^{0.03}_{0.06}$ & 114.21$\pm^{5.80}_{4.75}$ & 1.90$\pm^{0.08}_{0.10}$ & 52$\pm^{13}_{13}$ & 2.56$\pm^{0.22}_{0.20}$ & 4.0$\pm^{1.6}_{0.7}$ & 2.59$\pm^{0.19}_{0.22}$ & 550$\pm^{70}_{90}$  & pre \\ 
TYC\,56-284-1 & 15$\pm$3 & 40 & 6800$\pm$300 & 4.6$\pm$0.6 & 0.14$\pm^{0.20}_{0.13}$ & 91.28$\pm^{0.00}_{1.86}$ & 1.76$\pm^{0.04}_{0.01}$ & 6$\pm^{1}_{1}$ & 1.47$\pm^{0.06}_{0.06}$ & 13$\pm^{4}_{3}$ & 1.42$\pm^{0.10}_{0.05}$ & 2800$\pm^{200}_{900}$  & pre \\ 
TYC\,1235-186-1 & 35$\pm$3 & 50 & 6100$\pm$300 & 4.5$\pm$0.6 & 0.96$\pm^{0.54}_{0.34}$ & 54.41$\pm^{3.04}_{3.68}$ & 2.81$\pm^{0.19}_{0.16}$ & 10$\pm^{2}_{2}$ & 1.93$\pm^{0.43}_{0.32}$ & 5$\pm^{4}_{2}$ & 1.93$\pm^{0.15}_{0.46}$ & 800$\pm^{2000}_{200}$  & post \\ 
HIP\,17527 & 5$\pm$3 & 210 & 12500$\pm$400 & 4.1$\pm$0.5 & 0.33$\pm^{0.00}_{0.01}$ & 46.03$\pm^{0.85}_{0.65}$ & 3.00$\pm^{0.05}_{0.06}$ & 197$\pm^{23}_{23}$ & 3.47$\pm^{0.02}_{0.22}$ & 1.69$\pm^{0.24}_{0.05}$ & 3.45$\pm^{0.04}_{0.19}$ & 260$\pm^{42}_{9}$  & pre \\ 
HIP\,17543 & 2$\pm$4 & 60 & 12600$\pm$900 & 4.1$\pm$0.6 & 0.21$\pm^{0.03}_{0.03}$ & 64.35$\pm^{3.66}_{2.81}$ & 2.38$\pm^{0.10}_{0.14}$ & 130$\pm^{40}_{40}$ & 3.35$\pm^{0.19}_{0.39}$ & 2.1$\pm^{0.7}_{0.4}$ & 3.31$\pm^{0.23}_{0.35}$ & 300$\pm^{100}_{50}$  & pre \\ 
		 ... & ... & ... & ... & ... & ... & ... & ... & ... & ... & ...   \\
		\hline
	\end{tabular*}
	\begin{description}
      \item \textbf{Notes:} Only a portion of this table is presented here to show its content and description. Given its size, the full table is only available at the CDS. 
    
	\end{description}
\end{sidewaystable}
\clearpage
\normalsize

For the cases where more than one spectrum was obtained, we combined all the observations into a single `median spectrum' by computing the statistical median of all the spectra (previously normalized to the continuum) in order to achieve better signal to noise (SNR). Poor SNR observations (SNR$\lesssim$20) were discarded and not included in the median.

\section{Methods and Results}

\subsection{Estimation of Stellar Parameters}
\label{sec:stellarparams}

\subsubsection{Radial Velocity Measurement}
\label{sec:RV}

In order to find the best fitting synthetic spectral model for the spectral lines it is necessary to either shift the model's wavelength to match the radial velocity of the science spectra or to center the science spectra to the wavelength of rest to match the models. Either way, a measurement of radial velocity is needed (\citealt{Frasca2017}, \citealt{Iglesias2018}).

There is a marked difference between the spectra of mainly radiative, hotter stars and the spectra of cooler stars with a convective surface. On one hand there are the spectra of earlier types (early F, B, A and O) that have a radiative surface and, in general, fewer spectral lines that are, in addition, frequently rotationally broadened (specially in the case of A-type stars). On the other hand, the spectra of later types (late F, G, K, M) with a convective surface layer show overall plenty of spectral lines having narrow widths due to, for the most part, their lower rotational velocities. The boundary between `early' and `late' type stars is typically considered to be middle-F (F5), although this limit is not strict given that chromospheric emission lines have also been detected in stars earlier than F5 (\citealt{Mizusawa2012}, \citealt{Linsky2017}).

The standard procedure to measure radial velocities of stars typically consists of cross-correlating the spectra with a binary mask or a radial velocity standard star (e.g. \citealt{Baranne1979}, \citealt{Kurtz1998}, \citealt{Elliott2014}). This method can provide accurate radial velocity measurements for late type stars, but it can result in very large uncertainties for the early types. A simple method, such as measuring the centroid of strong lines gives more precise measurements of radial velocities for the case of B, A and early F type stars (e.g. \citealt{Iglesias2018}, \citealt{Torres2020}). Therefore, we divided our sample in "early" and "late" subsamples by visually inspecting the spectra and classifying according to the characteristics of a radiative or convective surface as described above.

\begin{table}
	\centering
	\caption{Adopted and explored values for the parameters of the Kurucz models.}
	\label{tab:synthe}
	\begin{tabular*}{\columnwidth}{l@{\extracolsep{\fill}}c} 
		\hline
		Parameter & Values \\
		\hline
		Turbulent Velocity & 2.0 km s$^{-1}$\\
		Additional Turbulence & 0.0 km s$^{-1}$ \\
		Opacity Threshold & 0.001  \\
		T$_{\rm eff}$ &  3500--20000 K, with $\Delta$=250 K \\
		$\log (g)$ & 3.5--5.0 dex, with $\Delta$=0.1 dex \\
		$\rm [Fe/H]$  & 0.0  \\
		$v$sin$i$  & 0--400 km s$^{-1}$, with $\Delta$=10 km s$^{-1}$ \\
		\hline
	\end{tabular*}
\end{table}

We measured radial velocities of the "late" group (73 stars) by means of the cross-correlation method using binary masks as stellar templates. For the "early" subsample (168 stars) we followed a procedure similar to \cite{Iglesias2018}. We fitted Lorentzian profiles to five photospheric Balmer lines in order to measure the centroid of these lines and estimated radial velocities and their uncertainties from the average and standard deviation among these lines measurements. We included H$\alpha$, H$\beta$, H$\gamma$, H$\delta$ and H$\zeta$ in the calculation. We did not include H$\epsilon$ since it is blended with the Ca\,{\sc ii}\,H line at 3968.47\AA. For the cases where strong emission was found in H$\alpha$, this line was excluded from the estimates given that it is not possible to obtain a good centering of the absorption profile. For those objects having more than one epoch we averaged the measurements from all the epochs and included the dispersion in the uncertainties by error propagation. The radial velocity measurements can be found in Table \ref{tab:params}.

\subsubsection{Effective Temperature, Surface Gravity and Projected Rotational Velocity}
\label{sec:Teffloggv}

We used Kurucz models \citep{Castelli1997} to estimate effective temperature (T$_{\rm eff}$), surface gravity ($\log (g)$) and projected rotational velocity ($v$sin$i$) for the sample. We computed synthetic models using the spectral synthesis codes \texttt{SYNTHE} and \texttt{ATLAS 9} \citep{Sbordone04}. We built a grid of models covering the UV-optical wavelengths with a spectral resolution similar to that of our VIS X-Shooter data (i.e. R=10000). We used the standard values for turbulent velocity, additional turbulence and opacity threshold offered by \texttt{ATLAS 9}, as shown in Table \ref{tab:synthe}. For the T$_{\rm eff}$ we explored a wide range of values ranging from 3500--20000 K in steps of 250 K, for the $\log (g)$ we covered a range from 3.5--5.0 dex in steps of 0.1 dex, and for the $v$sin$i$ we covered the range from 0--400 km s$^{-1}$ in steps of 10 km s$^{-1}$. These ranges should generously cover all the possible T$_{\rm eff}$, $\log (g)$ and $v$sin$i$ values in our sample. As in previous works where a similar procedure to estimate stellar parameters was employed (e.g. \citealt{Fairlamb2015}, \citealt{Wichittanakom2020}, \citealt{Vioque2022}), we have also adopted a metallicity value of $\rm [Fe/H]$=0. Although, as discussed in these works, the remaining parameters could be affected by the metallicity, on the other side, by fixing a standard metallicity value we avoid introducing further degeneracy on the models, since different combinations of parameters can produce similar models. We explored the metallicity dependence by fixing a different value of $\rm [Fe/H]$ for a few objects. We found that for small differences in metallicity with respect to solar, such as $\rm [Fe/H]$=+0.2, the average T$_{\rm eff}$ and $\log (g)$ values were within 1$\sigma$, and for larger differences, e.g. $\rm [Fe/H]$=-1.0, T$_{\rm eff}$ could vary up to 2$\sigma$. The estimation of $v$sin$i$, however, was not affected by changes in metallicity. A summary of the range of parameters and adopted values used for the models is given in Table \ref{tab:synthe}.

\begin{figure*}
	\includegraphics[width=\textwidth]{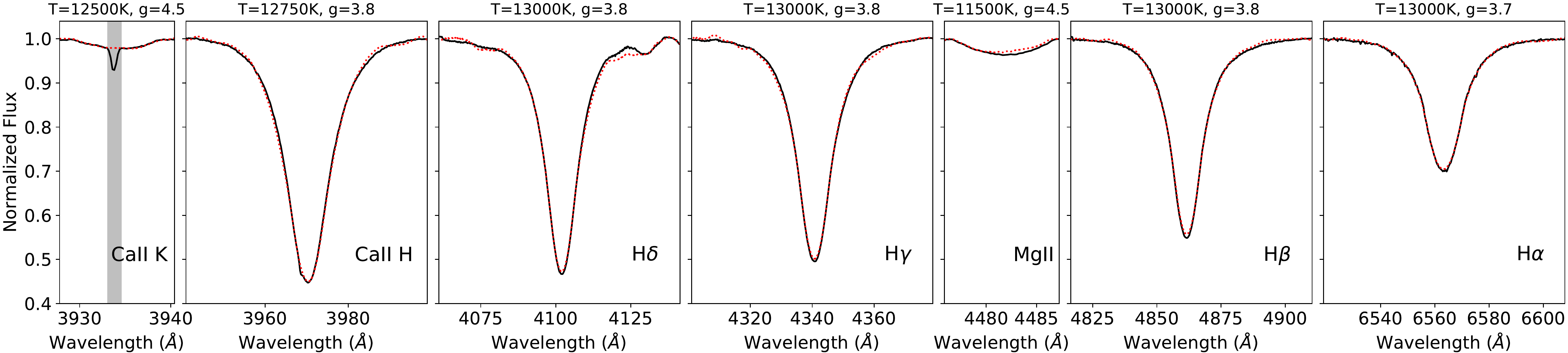}
	
	\vspace{0.3cm}
	
	\includegraphics[width=\textwidth]{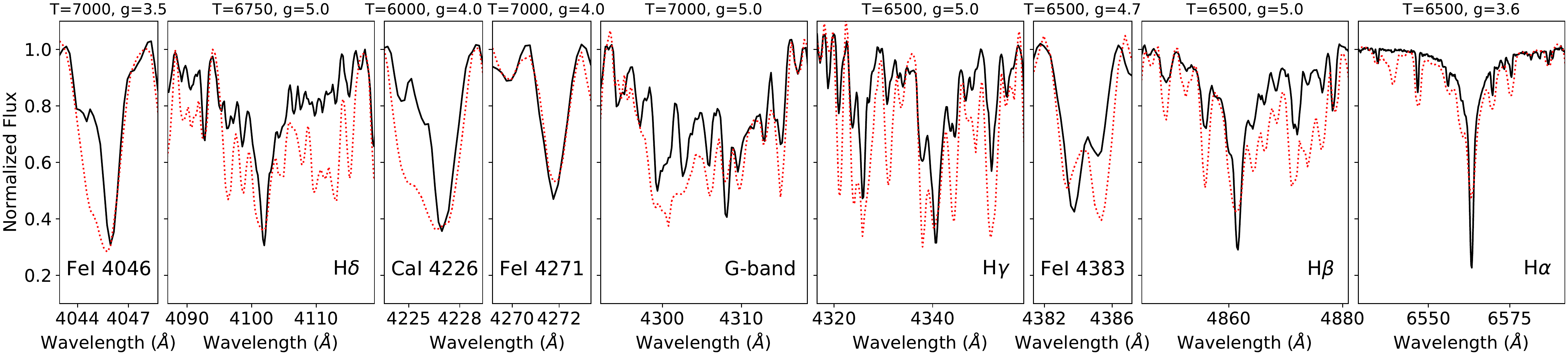}
    \caption{Examples of Kurucz model fitting for the ``early-type'' object HIP\,57451 (top) and for the ``late-type'' object HIP\,25309 (bottom). A $v$sin$i$ of 330 km s$^{-1}$ and 60 km s$^{-1}$ were determined for these early and late-type targets, respectively, and used as a fixed parameter for all the lines. The Ca\,{\sc ii}\,K line in HIP\,57451 presented interstellar absorption features which were excluded from the model fitting (shaded gray region). The best fitting T$_{\rm eff}$ and $\log (g)$ values for each line are indicated on top as 'T' and 'g', respectively. Kurucz models are shown in dotted red line and X-Shooter spectra in black.}
    \label{fig:model}
\end{figure*}

When comparing models to the science spectra, models were shifted in wavelength to match the radial velocities of the stars computed in the previous step. In addition, since the spectral resolution of the models is constant while the resolution of our observations differs for the three X-Shooter arms, we performed a Gaussian convolution to the models to match the specific spectral resolution of each wavelength range used for the model fit and resampled to match the spectral data points. Each spectral line was normalized to the continuum by fitting a portion of the continuum at each side of the absorption line. The model lines were normalized as well following the same procedure as for the science spectral lines for a good comparison. For the cases where the lines presented core emission features, the emission regions were masked and excluded from the fit as the models do not reproduce these emission features. The best fit models for each line were chosen by means of the minimum reduced $\chi^{2}$.

\begin{table}
	\centering
	\caption{Spectral lines used for the $v$sin$i$, T$_{\rm eff}$ and $\log (g)$ estimation.}
	\label{tab:lines}
	\begin{tabular*}{\columnwidth}{l@{\extracolsep{\fill}}cc} 
		\hline
		Element & Wavelength (\AA) & Usage \\
		\hline
		Ca\,{\sc ii}\,K & 3933.663 & `early' subset \\
		Ca\,{\sc ii}\,H & 3968.468 & `early' subset \\
		Fe\,{\sc i} &  4046.0620 & `late' subset \\
		H$\delta$  & 4101.7415 & `early' and `late' subsets \\
		Ca\,{\sc i} &  4226.73 & `late' subset \\
		Fe\,{\sc i} & 4271.7602 & `late' subset \\
		G-band & 4307.74 & `late' subset \\
		H$\gamma$  & 4340.471 & `early' and `late' subsets \\
		Fe\,{\sc i}  &  4383.5447 & `late' subset \\
		Fe\,{\sc i}  &  4404.7501 & only $v$sin$i$ \\
		Mg\,{\sc ii}  & 4481.126 & `early' subset and $v$sin$i$  \\
		H$\beta$  & 4861.3615 & `early' and `late' subsets \\
		H$\alpha$  & 6562.8518 & `early' and `late' subsets  \\
		\hline
	\end{tabular*}
\end{table}

We started by estimating the $v$sin$i$ in order to fix this value for all the lines used in the T$_{\rm eff}$ and $\log (g)$ assessment, since the effect of rotational broadening can strongly bias the T$_{\rm eff}$ estimation \citep{Royer2002}. We used the Mg\,{\sc ii} line at 4481.13 \AA\, to estimate the $v$sin$i$ value as this line is narrow and well isolated and thus, widely used for projected rotational velocity estimation. For the cases when this line was blended or too weak and did not allow a good fit, the Fe\,{\sc i} line at  4404.75 \AA\, was used instead as it offers similar characteristics. We searched for the best model fit for these lines within the grid of models described above and adopted the best fitting $v$sin$i$ value as a fixed parameter for the rest of the lines.

For the T$_{\rm eff}$ and $\log (g)$ estimation, we used subsets of the spectral lines as listed in Table \ref{tab:lines}. We split the sample in the two subsets defined previously for the radial velocity measurements; `early' and `late' subsamples, and used different sets of spectral lines for both cases since certain lines are weaker/stronger according to the spectral type. The spectral lines used for each subsample were carefully selected based on the studies of stellar atmospheres by \cite{GrayCorbally2009}, where it is discussed which lines are good T$_{\rm eff}$ and luminosity indicators for specific T$_{\rm eff}$ ranges. For the `early' subsample we used Ca\,{\sc ii}\,H \& K, Mg\,{\sc ii}, H$\alpha$, H$\beta$, H$\gamma$ and H$\delta$, and for the `late' subsample we used Ca\,{\sc i}, Fe\,{\sc i} at 4046, 4271 and 4383 \AA, H$\alpha$, H$\beta$, H$\gamma$, H$\delta$, and the G-band, as these were found to be the strongest lines and the best T$_{\rm eff}$ indicators overall for each subsample. The observed wavelengths of rest adopted for these lines were taken from the NIST\footnote{\url{https://physics.nist.gov/PhysRefData/ASD/lines_form.html}} Atomic Spectra Database. The lines used for each subsample and their wavelengths of rest are listed in Table \ref{tab:lines}.

T$_{\rm eff}$ and $\log (g)$ estimates were obtained by averaging the best fitting T$_{\rm eff}$ and $\log (g)$ of the lines used for the corresponding subset. As was done for the radial velocity estimation, for the cases where H$\alpha$ presented a strong emission, this line was excluded from the average as the emission dominates the underlying photospheric absorption line making the photospheric model fit less accurate. Errors were calculated as the quadratic sum of the standard deviation and the half of the step value used in the grid (shown in Table \ref{tab:synthe}). Estimations of $v$sin$i$, T$_{\rm eff}$ and $\log (g)$ for each object are presented in Table \ref{tab:params}. Examples of the Kurucz model fit for an early-type star and for a late-type star are shown in Figure \ref{fig:model}. 

\subsubsection{Visual extinction, distance and radius}
\label{sec:AvDR}

\begin{figure}
	\includegraphics[width=\columnwidth]{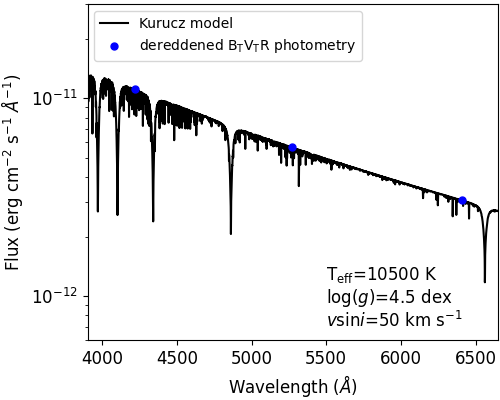}
	\includegraphics[width=\columnwidth]{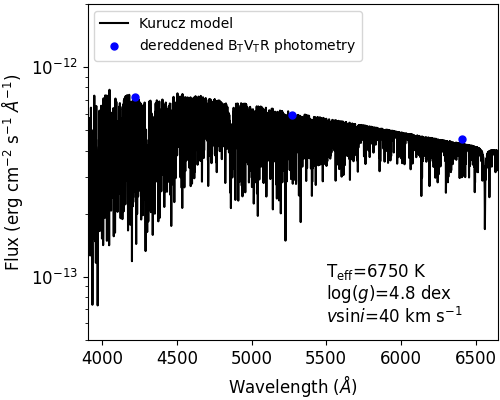}
    \caption{Example of photometry fitting for the ``early-type'' object HIP\,102719 (top), and for the ``late-type'' object TYC\,453-449-1. Stellar parameters used for each Kurucz model are indicated. The best fit between the model and BVR photometry for HIP\,102719 was found for a reddening value of $A_{V}$=0.52$\pm^{0.02}_{0.05}$ mag and a scaling ratio $D/R_{\star}$=85.53$\pm^{1.25}_{2.35}$ pc/R$_{\odot}$, and for TYC\,453-449-1 this was achieved for $A_{V}$=0.23$\pm^{0.20}_{0.23}$ mag and $D/R_{\star}$=126.50$\pm^{3.36}_{2.02}$ pc/R$_{\odot}$.}
    \label{fig:photofit}
\end{figure}

After obtaining the T$_{\rm eff}$, $\log (g)$ and $v$sin$i$ estimates, we used a Kurucz model spectrum with these values covering the full wavelength range provided by \texttt{SYNTHE} (i.e. 100-10,000 nm) in combination with photometric measurements to estimate the stellar radius $R_{\star}$ and visual extinction $A_{V}$. We followed a similar procedure to \cite{Fairlamb2015} and \cite{Wichittanakom2020} to estimate $R_{\star}$ and $A_{V}$, however, only B$_{\rm T}$V$_{\rm T}$R magnitudes were used in this case given that I magnitudes were not available for the full sample, and, for few particular cases, only B$_{\rm T}$ and V$_{\rm T}$ magnitudes were used when R was not available. Basically, the flux density $f$ of the Kurucz model and the (derredened) observed photometric fluxes are scaled by the square ratio of distance to the star and its radius ($(D/R_{\star})^{2}$). We first converted the B$_{\rm T}$V$_{\rm T}$R magnitudes to fluxes using the zero magnitude absolute flux densities from \cite{Bessell1998}. Then, we estimated the B$_{\rm T}$V$_{\rm T}$R fluxes from the model spectra from the convolution with the transmission curve of the B$_{\rm T}$V$_{\rm T}$R passbands.  We obtained the best fitting $A_{V}$ by varying $A_{V}$ in steps of 0.01 magnitude and finding the best alignment of the model with the dereddened photometry. Then, the $D/R_{\star}$ value was obtained by scaling with the dereddened V$_{\rm T}$-band flux. The observed photometry was dereddened each time using the extinction values from \cite{Cardelli1989} and adopting the standard value of total to selective extinction parameter $R_{V}$ = 3.1. The uncertainties in the $A_{V}$ and $D/R_{\star}$ ratio were estimated by following the same procedure but using the Kurucz model spectra corresponding to the upper and lower T$_{\rm eff}$ values (T$_{\rm eff}\pm uncertainty$) and finding the resulting lower and upper $A_{V}$ and $D/R_{\star}$ values. Finally, we used (photogeometric) distances obtained from the Gaia DR3 catalogue \citep{Gaia2022k,Bailer-Jones2021} to obtain $R_{\star}$ estimates from the $D/R_{\star}$ ratio. The uncertainties in the $R_{\star}$ values were obtained by error propagation of the corresponding distances and $D/R_{\star}$ uncertainties. Distances are shown in Table \ref{tab:liter} and estimates of $A_{V}$, $D/R_{\star}$ and $R_{\star}$ are listed in Table \ref{tab:params}. Examples of the photometry fitting for an early and a late-type star are illustrated in Figure \ref{fig:photofit}.

\begin{figure*}
	\includegraphics[width=0.97\textwidth]{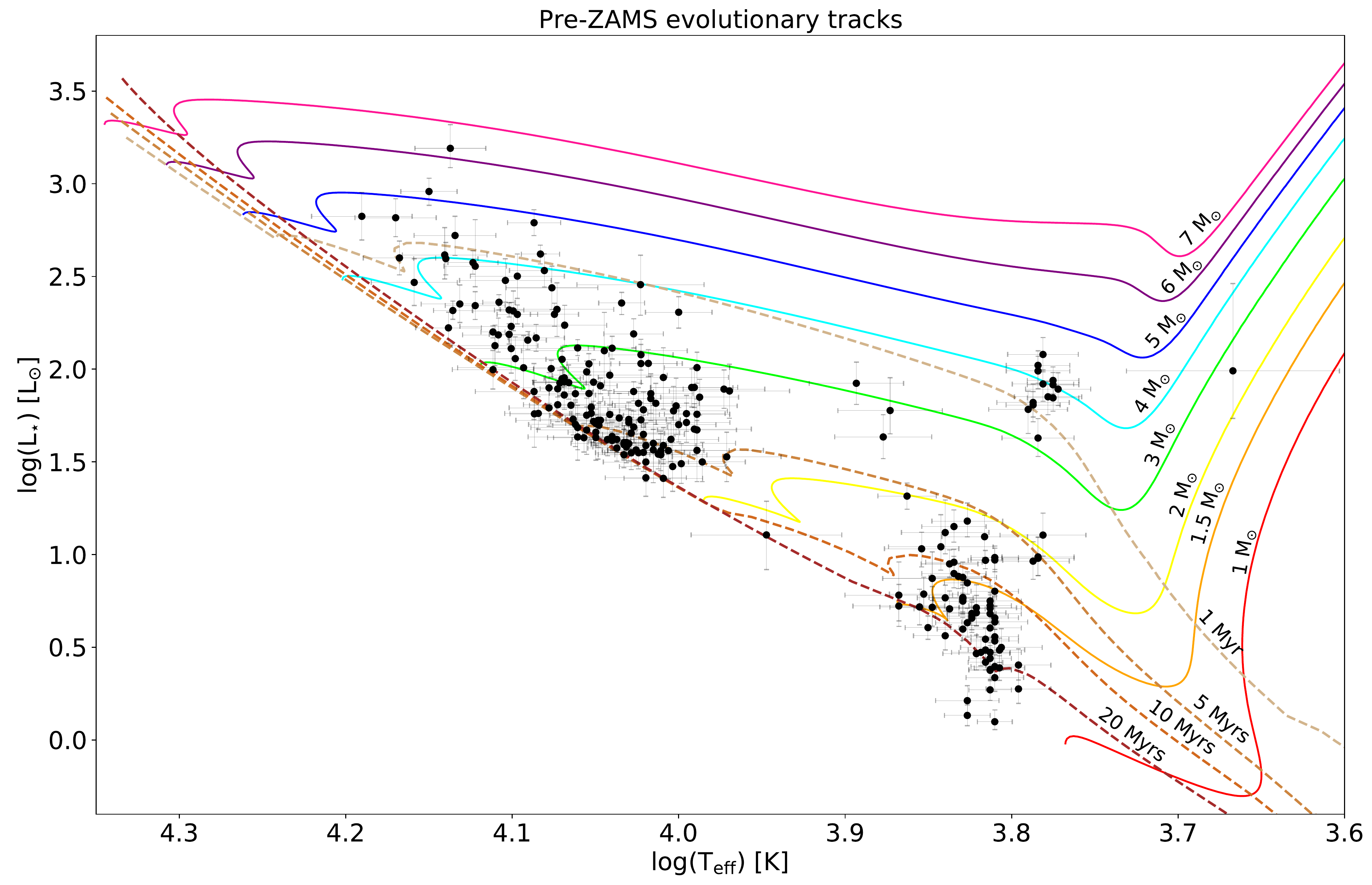}
	\includegraphics[width=0.97\textwidth]{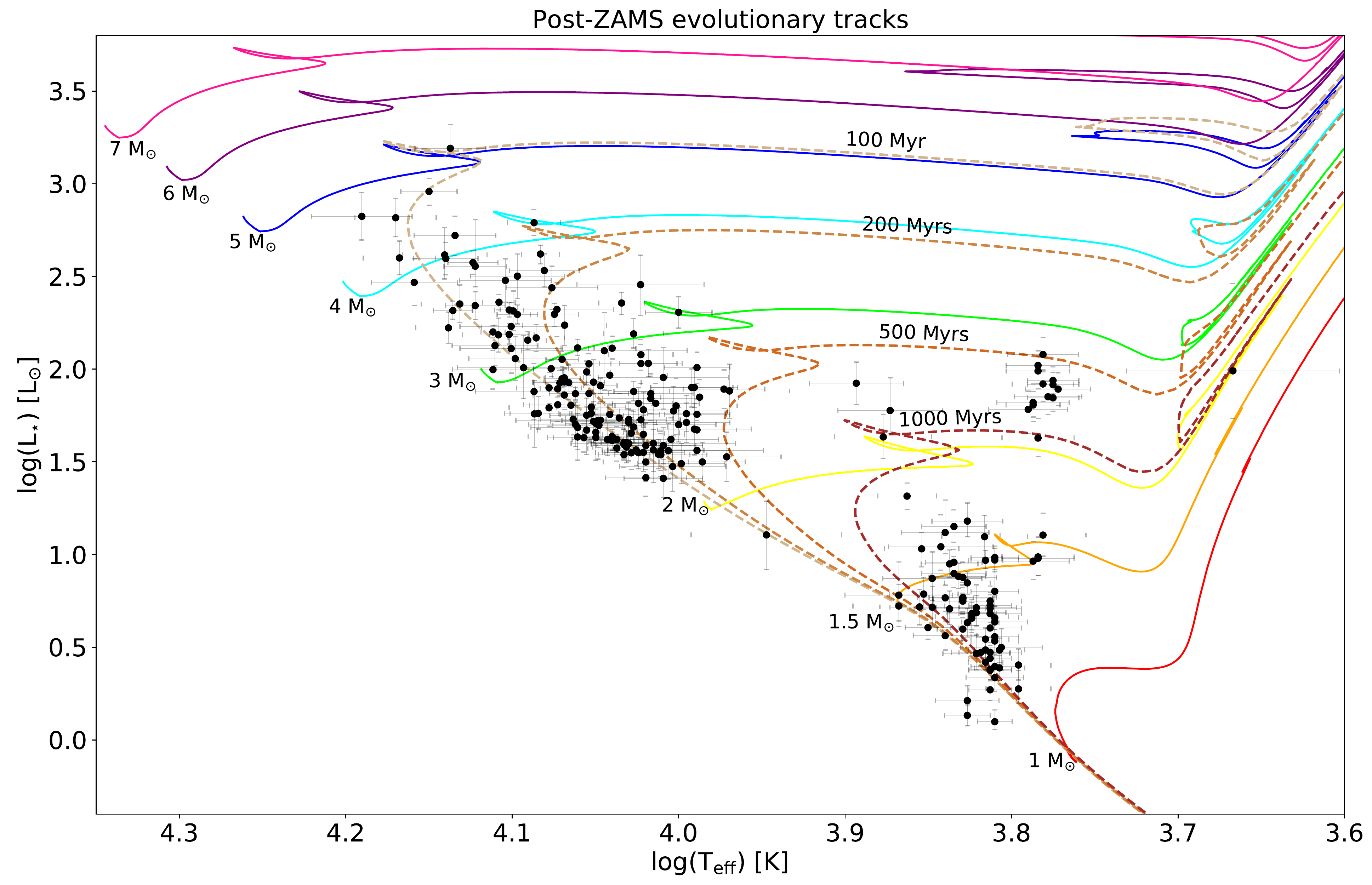}
    \caption{Full sample of 241 IMS candidates placed in the HR-diagram. \textbf{Top:} Pre-ZAMS evolutionary tracks with initial masses from 1 to 7 M$_{\odot}$ are shown in solid lines and isochrones of 1, 5, 10, and 20 Myrs in dashed lines. \textbf{Bottom:} Post-ZAMS evolutionary tracks with initial masses from 1 to 7 M$_{\odot}$ are shown in solid lines and isochrones of 100, 200, 500, and 1000 Myrs in dashed lines. Evolutionary tracks and isochrones are from \protect\cite{Bressan2012}.}
    \label{fig:HRD}
\end{figure*}

\subsubsection{Mass, Luminosity and Age}
\label{sec:MLumAge}

Luminosities were calculated using the Stefan-Boltzmann law: $L_{\star}=4\pi$R$^{2}_{\star}\sigma$T$^{4}_{\rm eff}$. Having luminosities and T$_{\rm eff}$ values, the stars were placed in the HR diagram, as shown in Fig.\ref{fig:HRD}. Initially, a set of stellar tracks and isochrones from \texttt{PARSEC} \citep{Bressan2012} for masses between 1 and 8 M$_{\odot}$, solar metallicity Z=0.0152 (same as the metallicity adopted for the Kurucz models), and ages from 0 to 100 Myr were used to estimate masses and ages by interpolating between the closest values to the stellar tracks. We also estimated the masses and ages using the evolutionary tracks from \cite{Siess2000} using masses between 1 and 7 M$_{\odot}$ (the maximum available), solar metallicity, and the same age range.  We compared our results for both sets of evolutionary tracks and we found that ages were in agreement within the first $\sim$20 Myrs and then started to diverge significantly, while masses were in good agreement overall. This is likely because around this age the intermediate mass stars reach the zero age main sequence (ZAMS), then their evolutionary tracks start to decrease in T$_{\rm eff}$ and overlap with the younger ages producing ambiguity in the results. Therefore, we decided to split the age of the stellar tracks at the turning point for each mass and obtain both pre and post ZAMS age values, as shown in Figure \ref{fig:HRD}. Both age values are presented in Table \ref{tab:params} along with the masses and luminosities. Given that our sample has been selected under criteria of being likely pre-main-sequence objects and having IR-excesses, we initially adopt the pre-main-sequence age and mass values as the most probable ones. Later, in Section \ref{sec:Ages} we analyse in more detail those objects that might be more evolved, and re-assign their masses and ages using the post-ZAMS stellar tracks. Finally, we chose to use the mass and age estimations based on the \texttt{PARSEC} stellar tracks rather than those of \cite{Siess2000} because they offered a finer and slightly more extended (in terms of mass) grid of stellar tracks, allowing a more precise interpolation.

\begin{figure}
	\includegraphics[width=\columnwidth]{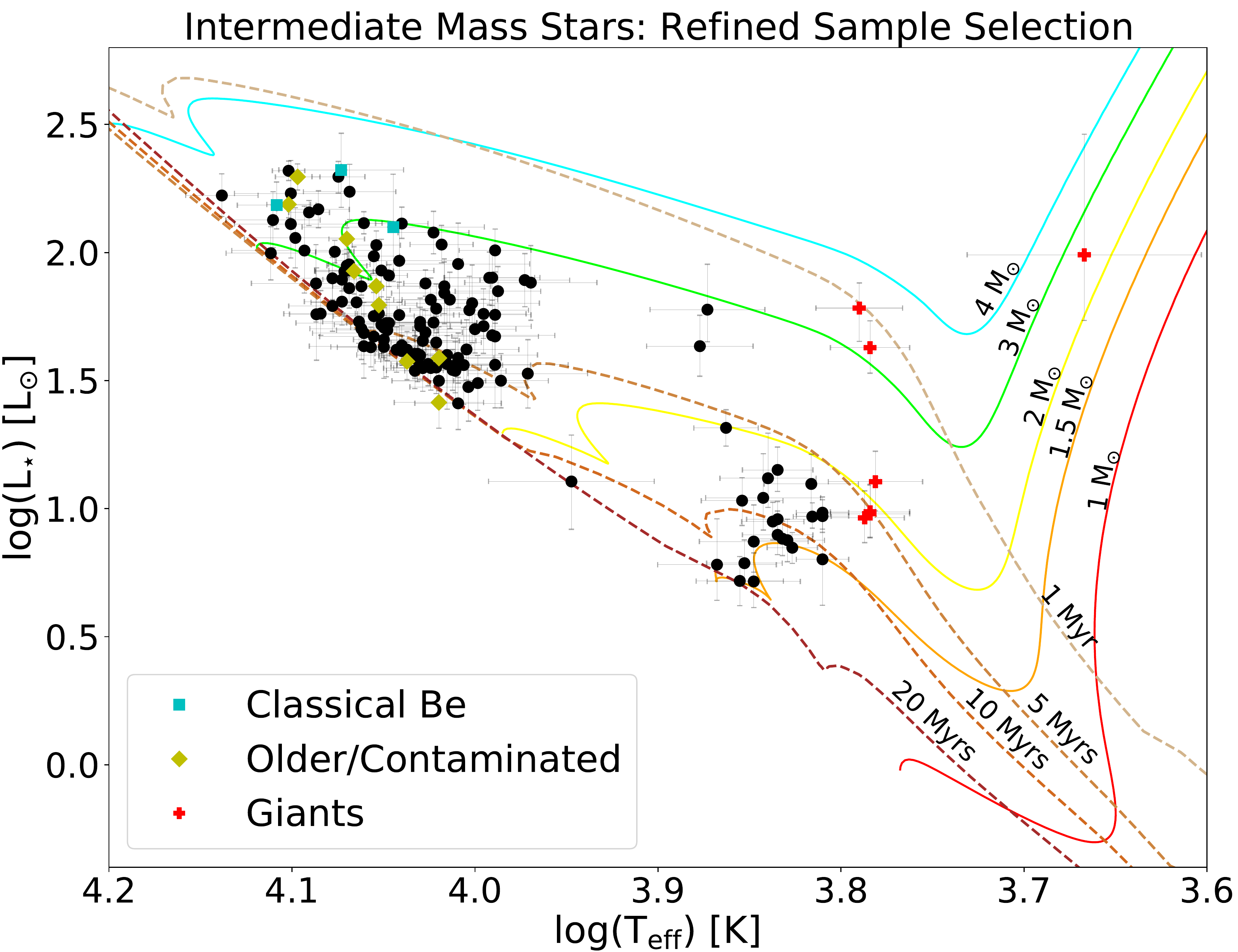}
    \caption{Selection of IMS within the range $1.5 \leq M_{\star}$/$M_{\odot} \leq 3.5$ and within $\leq$ 300 pc. Contaminants discarded from this selection due to being more evolved are highlighted in red crosses (giants), yellow diamonds (older/contaminated), and cyan squares (classical Be stars). Pre-main-sequence evolutionary tracks with initial masses from 1.5 to 4 M$_{\odot}$ are shown in solid lines and isochrones of 1, 5, 10, and 20 Myrs are plotted in dashed lines. }
    \label{fig:Selection}
\end{figure}

\subsection{Sample Refinement}
\label{sec:refined}

One of the main purposes of our survey is to study how circumstellar discs evolve around intermediate mass stars. In particular, we are interested in how fast the inner disc regions get cleared, and how they compare to low mass stars. 
Using the calculated basic stellar parameters and Gaia DR3 distances the sample is refined by selecting those with masses within the range $1.5 \leq M_{\star}$/$M_{\odot} \leq 3.5$ and with distances $\leq$ 300 pc for further analysis of their IR excesses. As the distances were updated from the initial selection made using TGAS, now 95\% of the objects had distances $\leq$ 300 pc when using Gaia DR3. We found that 
154$\pm^{29}_{23}$  (63.9$\pm^{12.0}_{9.5}$\%) of our candidates fall within the refined selection. Lower and upper uncertainties are derived considering lower and upper uncertainties in masses and distances simultaneously, thus estimating the minimum and maximum number of objects that might fall within the selection. The placement in the HR diagram for the stars in the refined sample is shown in Figure \ref{fig:Selection}.

\subsection{Infra-Red excess}
\label{sec:IRexs}

\cite{Wyatt2015} suggest that a helpful classification of circumstellar discs could be based on the fractional IR-excess $f_{\lambda}$, which is defined as the ratio of the total flux measured from a system to that of the stellar flux at a given IR wavelength $\lambda$, i.e.: $f_{\lambda}$=$F(\lambda)/F_{\star}(\lambda)$. Once obtained the reddening $A_{V}$ and the flux ratio $D/R_{\star}$, it is possible to estimate the fractional $f_{\lambda}$  above the stellar flux by using again the Kurucz model spectra as stellar templates. We have used $K_{s}$-band photometry from 2MASS survey \citep{Skrutskie2006} and W1, W2, W3 and W4 bands photometry from WISE survey \citep{Cutri2014} to estimate fractional IR-excesses $f_{\lambda}$ at 2.2, 3.4, 4.6, 12 and 22 $\rm{\mu}$m, respectively. Since the Kurucz models only span a wavelength range between 0.1 and 10 $\rm{\mu}$m, we extended the wavelength coverage up to 25 $\rm{\mu}$m by fitting a Planck function to the models in order to cover the W3 and W4 bands. We consider a star to have IR-excess at a given $\lambda$ when $f_{\lambda}$ is `firmly' $>$ 1 (i.e. 1$\sigma$ above 1). The mentioned photometry was derredened using the estimated $A_{V}$ values and following the wavelength dependence as defined in \cite{Prato2003}. Uncertainties were derived by error propagation of the photometry, $A_{V}$, $D/R_{\star}$, and flux. The $f_{\lambda}$ values are presented in Table \ref{tab:Fractexs}.

\subsection{Assessment of Ages}
\label{sec:Ages}

As described in Section \ref{sec:MLumAge}, age estimates were adopted from the pre-ZAMS stellar tracks in order to avoid ambiguity in the results but this, of course, biases the age estimation of possibly more evolved stars. Therefore, in this section we study in more detail those objects that might be older and more likely post-main-sequence. We start with the emission-line objects; we have found 22 objects in our preliminary sample with emission in their Balmer lines. They could either be Herbig Ae/Be stars or classical Be (CBe) stars, both presenting similar emission lines and usually hard to distinguish based on their spectra and their position in the HR diagram. However, Herbig Ae/Be are young pre-main-sequence (pre-MS) stars \citep{Hillenbrand1992}, while CBe stars are considered to be more evolved objects \citep{Rivinius2013}. A search of the literature finds that six of them were classified as Herbig Ae/Be stars, 13 as CBe stars, based on their emission lines and presence/lack of IR-excess, and three of them have not been yet classified to the best of our knowledge. In addition, we analyzed their estimated masses and fractional excesses and compared them based on their classification in the literature. We found that, overall, CBe stars were more massive than the Herbig ones, and the Herbig Ae/Be stars had much larger fractional excesses than the CBe stars. Considering this, the three previously unclassified objects; HIP\,23201, HIP\,77289, and HIP\,92364 were more consistent with the characteristics of the CBe stars, and thus we assign them this classification. The six accreting Herbig Ae/Be stars are analysed in more detail in Appendix \ref{sec:accretion}.

The estimated ages for the Herbig Ae/Be stars are consistent with them being pre-MS, but the initial ages estimated for the CBe stars ($\lesssim$2 Myrs) are unlikely to be so young. Although some studies have found CBe stars in clusters of age 5-8 Myrs \citep{Fabregat_Torrejon2000}, it has been suggested that, overall, CBe stars are in the second half of their main sequence lifetime and tend to be older than 10 Myrs (\citealt{Wisniewski_Bjorkman2006}, \citealt{Zorec2005}). Therefore, we opt for re-assigning the ages for the CBe stars to that estimated from the post-ZAMS stellar tracks, as described in Section \ref{sec:MLumAge}. Three out of the 16 CBe stars fall within the refined selection of IMSs and since our study is devoted to study disc evolution in young stars, these were excluded from the sample. The 13 remaining CBes were more massive than 3.5 M$_{\odot}$.

Then we assess the presence of giant contaminants in the sample. Red giants stars share similar locations in the HR diagram as young pre-main-sequence stars but they are much more evolved. We examined the possible giant contaminants in three ways: by measuring the Li\,{\sc i} line at 6708\AA, by analysing gravity-sensitive and T$_{\rm eff}$-sensitive spectral indices, and by looking at their IR-excesses and shape of their SEDs. Lithium depletion can be used as an age indicator for F, G, and K-type stars and the Li\,{\sc i} line at $\sim$6708\AA~ is a good abundance tracer (\citealt{Vican2012}, \citealt{Soderblom2014}). We measured the equivalent width (EW) of the mentioned Li\,{\sc i} line for the late-type stars in our sample and classified those with EW$<$0.1\AA\, as Li-poor and thus likely older. In addition, we calculated the gravity-sensitive and T$_{\rm eff}$-sensitive spectral indices $\gamma_{1}$ and $\beta_{t}$, respectively, following \cite{Damiani2014}, and analyzed their placement in the $\gamma_{1}$ vs. $\beta_{t}$ diagram. These spectral indices basically measure the strength of a set of gravity- (or T$_{\rm eff}$) sensitive lines. As giant stars have lower gravities than pre- and MS stars their low-gravity-dependant lines are stronger. In Figure 28 of \cite{Damiani2014} they show how $\gamma_{1}$ increases as $\beta_{t}$ decreases for giant stars, while pre-MS stars remain closer to $\gamma_{1}$=1. In comparison, some of the stars in our sample show clear signs of being giant contaminants going upwards in the $\gamma_{1}$ vs. $\beta_{t}$ diagram. After this analysis, those Li-poor objects with spectral indices consistent with giant stars and a lack of IR-excess were considered to be evolved contaminants and their ages were re-assigned to post-MS. Seven of these older contaminants were found within the refined selection of IMSs (as described in Seccion \ref{sec:refined}) and were removed from the sample. 

We also checked whether our targets belonged to stellar associations or star forming regions. We used the BANYAN $\Sigma$ tool (the Multivariate Bayesian Algorithm to identify members of Young Associations; \citealt{Gagne2018}) which calculates the probability of belonging to stellar associations within 150 pc of the Sun. We found 22 objects (15\% of our targets) with $\geq$ 99\% probability of belonging to an association. These associations have age estimations and some are older than 100Myrs (see Table 1 in \citealt{Gagne2018} and references therein). Therefore, we discard any (member) star with an age that is not compatible with the age of the association. We found eight stars that belong to associations older than $\sim$30Myrs (incompatible with their pre-MS stellar ages) and thus more likely to be post-MS. The remaining 14 stars had ages in agreement with the young age of their respective associations.
Finally, we visually inspected the AllWISE colored infrared images \citep{Wright2010} and the DSS colored optical images \citep{York2000} for each of our targets, looking for possible contamination that might lead to a 'false' IR-excess. We found six cases where the star was affected by contamination of reflection nebulae and therefore in a dust-rich region. Since this interstellar dust might affect the IR-excess measurement, we discard these stars to be cautious. It is worth mentioning, that five out of these six stars were among those eight stars we found to belong to older stellar associations. Furthermore, these five stars had the largest age difference with respect to their association. It is possible that the contamination might have had an impact in the estimation of the stellar parameters (the extinction, in particular, and as a consequence the luminosity) affecting the age determination. More details regarding these stars can be found in the Appendix \ref{sec:comments}.

In total, 19 objects (3 CBe, 7 giants, and 9 stars in older associations and/or with contamination) were excluded from the refined selection, leaving the sample  with a new total of 135 objects within 1.5--3.5$M_{\odot}$ and $<$300pc. Those contaminants excluded from this final selection are highlighted in the HR diagram (see Figure \ref{fig:Selection}). A list of these objects is given in the Appendix \ref{sec:comments}.


\begin{figure*}
	\includegraphics[width=\textwidth]{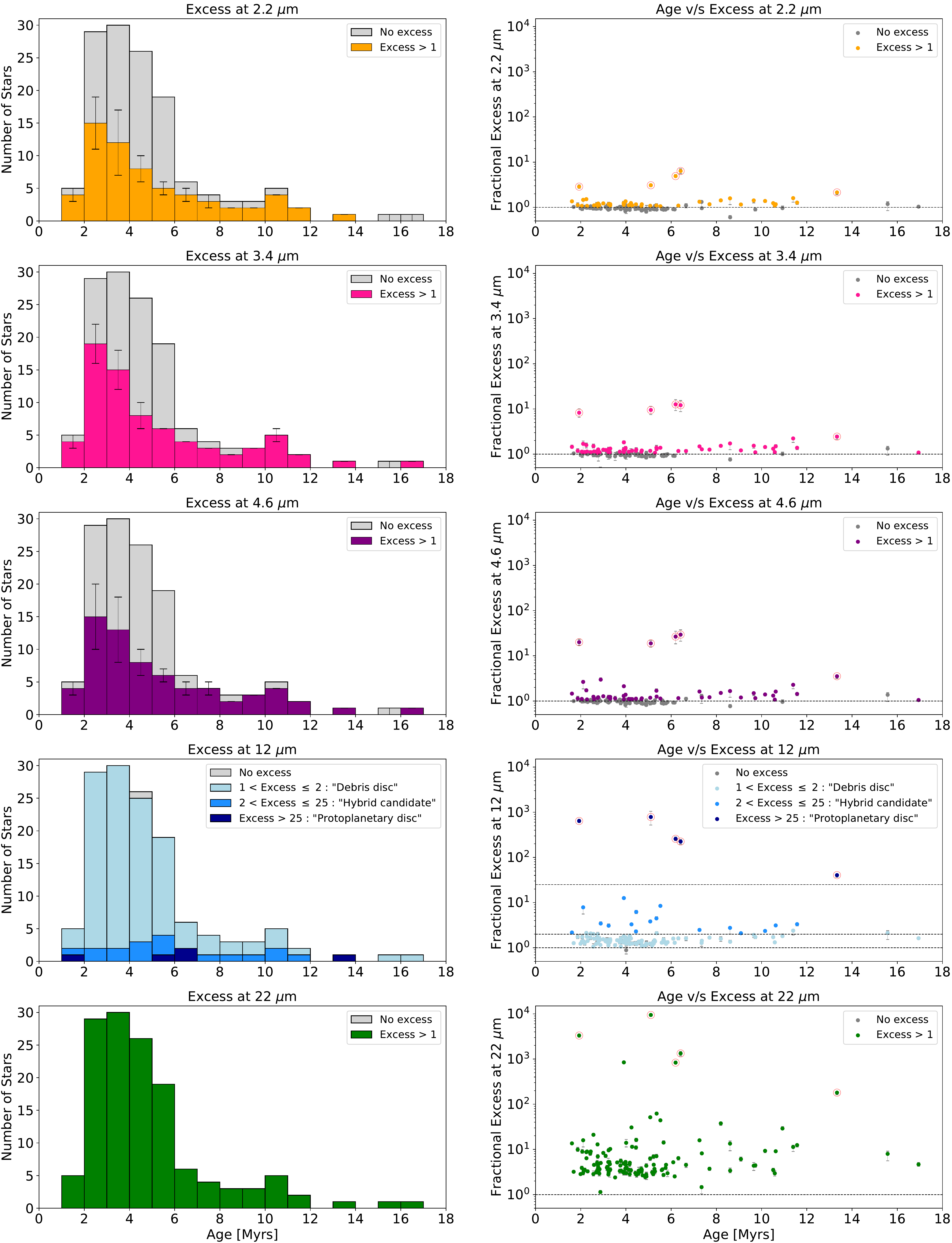}
   \caption{Left: Histograms of fractional excesses. Right: The corresponding age v/s excess values. From top to bottom: excess in K$_{s}$, W1, W2, W3 and W4 bands. Typical age uncertainties are in the order of $\sim$2 Myrs. Protoplanetary discs are highlighted in red circles. Error bars might be smaller than markers.}
   \label{fig:hist}
\end{figure*}

\section{Analysis and Discussion}
\label{sec:analysis_discu}

\subsection{Disc Classification}
\label{sec:Classif}

We classify our sample into circumstellar disc evolutionary stages. This classification is motivated by \cite{Wyatt2015} (W15), where the evolution from protoplanetary to debris disc was studied based on observations of fractional excesses. Figure 1 in W15 shows age versus fractional excesses at 12, 22, and 70 $\mu$m for a selection of known protoplanetary and debris discs around A-type stars (which fall into the category of intermediate mass stars). 
When looking at the fractional excesses ($R_{\lambda}$ in their paper) at 12 $\mu$m in Figure 1 of W15 (left top panel), there is a clear distinction between the distribution of protoplanetary and debris discs. Indeed, all known debris discs in that study have fractional excesses 1\,$< f_{12 \mu m} \leq$\,2, while protoplanetary discs range from values of $f_{12 \mu m}$ above 25 up to over 1000. 

\begin{table*}
	\centering
	\normalsize
	\caption{Fractional excesses and disc classification based on $f_{12 \mu m}$.}
	\label{tab:Fractexs}
			\begin{tabular*}{\textwidth}{l@{\extracolsep{\fill}}cccccc} 
		\hline
		Name & $f_{2.2\mu m}$  & $f_{3.4 \mu m}$ & $f_{4.6 \mu m}$ & $f_{12 \mu m}$  &  $f_{22 \mu m}$ & Disc classif. \\
		\hline
HIP\,13910 & 0.92$\pm^{0.09}_{0.08}$ & 0.99$\pm^{0.10}_{0.09}$ & 0.96$\pm^{0.10}_{0.08}$ & 1.41$\pm^{0.14}_{0.12}$ & 4.55$\pm^{0.52}_{0.45}$ & Debris \\ 
HIP\,17543 & 0.94$\pm^{0.11}_{0.08}$ & 0.95$\pm^{0.13}_{0.11}$ & 0.98$\pm^{0.11}_{0.09}$ & 1.74$\pm^{0.20}_{0.15}$ & 8.96$\pm^{1.04}_{0.80}$ & Debris \\ 
HIP\,21024 & 0.92$\pm^{0.06}_{0.06}$ & 0.93$\pm^{0.06}_{0.06}$ & 0.89$\pm^{0.06}_{0.05}$ & 1.22$\pm^{0.07}_{0.07}$ & 4.56$\pm^{0.31}_{0.30}$ & Debris \\ 
HIP\,22402 & 1.16$\pm^{0.13}_{0.14}$ & 1.11$\pm^{0.13}_{0.15}$ & 1.10$\pm^{0.12}_{0.13}$ & 1.60$\pm^{0.17}_{0.20}$ & 3.28$\pm^{0.38}_{0.43}$ & Debris \\ 
HIP\,23633 & 1.04$\pm^{0.06}_{0.05}$ & 1.06$\pm^{0.08}_{0.07}$ & 1.04$\pm^{0.06}_{0.05}$ & 3.30$\pm^{0.19}_{0.15}$ & 30.65$\pm^{1.75}_{1.38}$ & Hybrid \\ 
HIP\,24092 & 0.88$\pm^{0.06}_{0.05}$ & 0.92$\pm^{0.07}_{0.06}$ & 0.89$\pm^{0.06}_{0.05}$ & 1.20$\pm^{0.09}_{0.07}$ & 2.83$\pm^{0.32}_{0.30}$ & Debris \\ 
HIP\,56379 & 3.09$\pm^{0.18}_{0.22}$ & 9.50$\pm^{1.94}_{1.98}$ & 18.85$\pm^{3.09}_{3.18}$ & 784$\pm^{263}_{265}$ & 9448$\pm^{514}_{633}$ & Protoplanetary \\ 
TYC\,6487-537-1 & 1.09$\pm^{0.06}_{0.05}$ & 1.11$\pm^{0.06}_{0.04}$ & 1.07$\pm^{0.06}_{0.04}$ & 1.32$\pm^{0.08}_{0.05}$ & 2.95$\pm^{0.39}_{0.37}$ & Debris \\ 
HIP\,25453 & 0.92$\pm^{0.05}_{0.06}$ & 0.92$\pm^{0.08}_{0.09}$ & 0.93$\pm^{0.05}_{0.06}$ & 1.24$\pm^{0.07}_{0.08}$ & 4.58$\pm^{0.28}_{0.31}$ & Debris \\ 
HIP\,25763 & 0.90$\pm^{0.06}_{0.07}$ & 1.10$\pm^{0.07}_{0.08}$ & 1.16$\pm^{0.07}_{0.09}$ & 1.73$\pm^{0.10}_{0.13}$ & 4.44$\pm^{0.34}_{0.41}$ & Debris \\ 
		 ... & ... & ... & ... & ... & ... & ... \\
		\hline
	\end{tabular*}
	\begin{description}
      \item \textbf{Notes:} Only a portion of this table is presented here to show its content. Given its size, the full table is only available at the CDS. 
	\end{description}
\end{table*}

We, therefore, classify as ``debris'' discs those systems with fractional excess 1\,$< f_{12 \mu m} \leq$\,2 : 112 objects, as ``hybrid candidates'' those systems with 2\,$< f_{12 \mu m} \leq$\,25 : 17 objects, and as ``protoplanetary'' those with $f_{12 \mu m} >$\,25 : 5 objects. One object was found to have no excess at 12 $\mu$m. The different categories along with their ages and $f_{12 \mu m}$ are shown in Figure\,\ref{fig:hist}, right, 4$^{\rm th}$ row. We note that a number of our sources fall between the debris and protoplanetary categories, populating a desert where the only previously known object was HD\,141569, as shown in Fig. 1 of W15. We will further discuss this group of objects in Section \ref{sec:hybrids}.

\begin{figure}
	\includegraphics[width=\columnwidth]{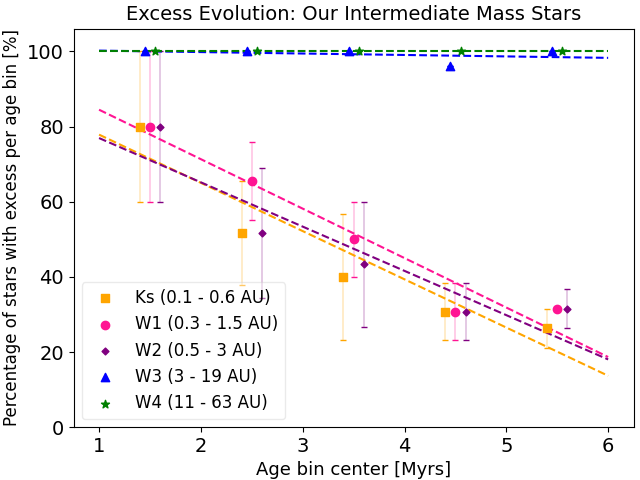}
    \caption{Percentage of stars with excess per age bin $\leq$ 6 Myrs at 2.2, 3.4, 4.6, 12 and 22 $\mu$m (K$_s$, W1, W2, W3, and W4 bands, respectively). The approximate location of the excess with respect to the star traced by each band is indicated. A small shift has been added to the horizontal position to avoid overlap. True age bin center is that of pink circles. 117 objects are included in this figure.}
    \label{fig:fraction}
\end{figure}

\begin{figure}
	\includegraphics[width=\columnwidth]{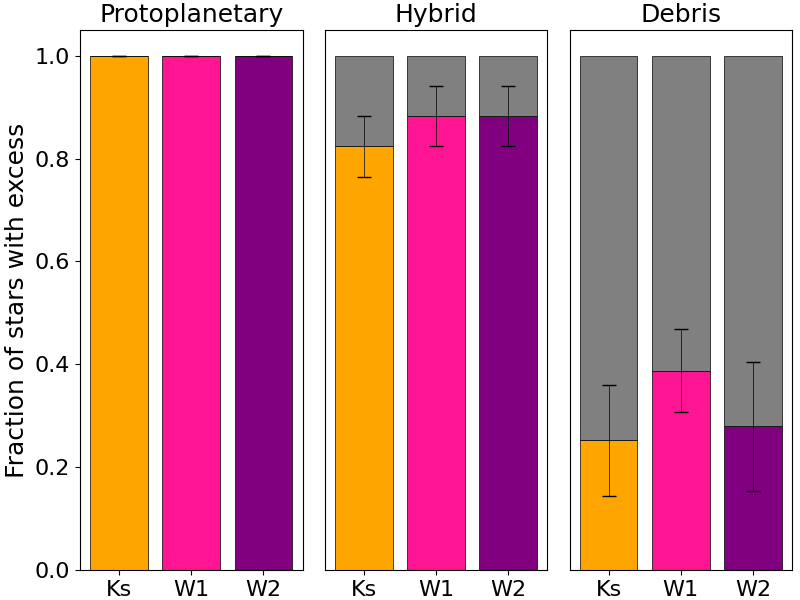}
    \caption{Comparison of excess fractions for $\lambda<$10$\mu$m for each disc classification based on fractional excess at 12$\mu$m. }
    \label{fig:discEvol}
\end{figure}

\begin{figure}
	\includegraphics[width=\columnwidth]{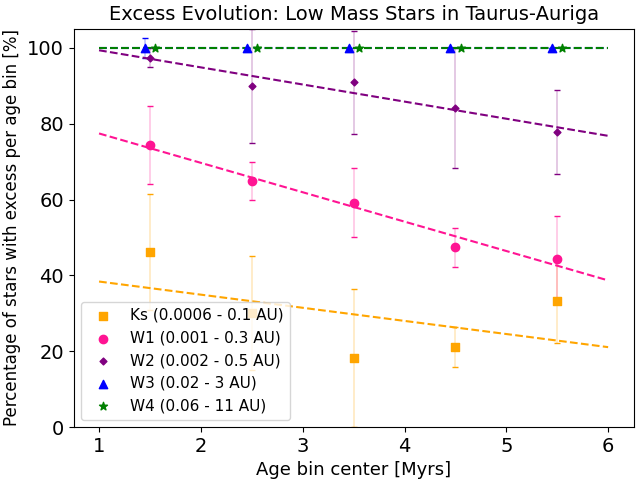}\\
	
	\includegraphics[width=\columnwidth]{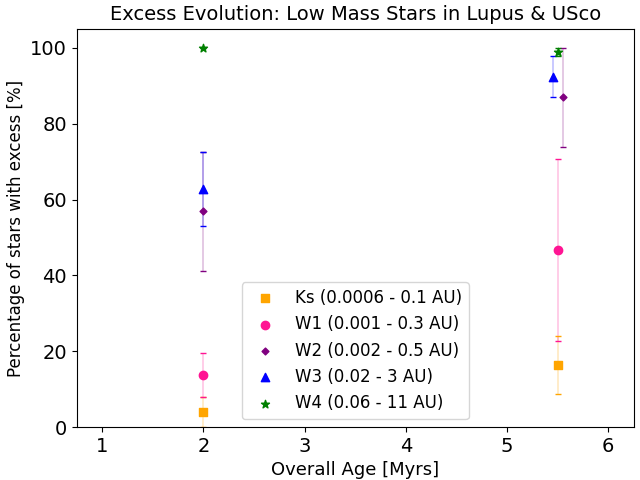}
    \caption{Same as Fig. \ref{fig:fraction} but for stars with masses $<$ 1.5 $M_{\odot}$. \textbf{Top:} Stars with ages $\leq$ 6 Myrs from the Taurus-Auriga star forming region. \textbf{Bottom:} Stars aged between 1--3 Myrs from Lupus cloud 3 (shown at 2 Myrs) and stars aged 5--11 Myrs from Upper Scorpius (placed at the 5--6 Myrs bin for comparison). Sample of Taurus-Auriga taken from \protect\cite{Andrews2013}, including 123 objects in the figure. Sample of Lupus taken from \protect\cite{Comeron2009}, including 42 objects, and sample from Upper Scorpius taken from \protect\cite{Barenfeld2016}, including 92 objects.} 
    \label{fig:TauAur}
\end{figure}


\subsection{IR-excess statistics}
\label{sec:statistics}

We studied the overall age versus near-IR excess trends across the different wavelengths by dividing our sample selection in age bins of 1 Myr, as shown in Figure\,\ref{fig:hist} (left). The corresponding fractional excess $f_{\lambda}$ values calculated for each wavelength are shown at the right side of Figure\,\ref{fig:hist}, along with the ages. The excess level classification at 12 $\mu$m is shown as well (fourth panel from top to bottom). All the stars have excess at 22 $\mu$m as this was an initial selection criterion for the sample.

In order to have a better picture of the excess evolution at different wavelengths, we computed the fraction of stars with excess per age bin for each wavelength. Since the number of IMSs in the sample drops dramatically for ages $>$6 Myrs and the number of stars per bin after this age are too few to obtain reliable statistics, we will limit this analysis up to ages $<$6 Myrs. 
The results are shown in Figure \ref{fig:fraction}, where the approximate location of the excess with respect to the star traced by each band is indicated, considering stars of 1.5$M_{\odot}$ and 3.5$M_{\odot}$ as lower and upper limits for the distance range. The approximate locations were derived by using Wien's displacement law to estimate the temperature observed for each band, and then using Stefan–Boltzmann law to calculate the distance from the star where dust would heat to such a temperature, adopting typical values for T$_{\rm eff}$ $\sim$7000K -- 13000K, and for R$_{\star}$ $\sim$1.7 -- 3 R$_{\odot}$, for the range.  We can see similar percentages of stars with excess for those wavelengths $<$10$\mu$m (K$_{s}$, W1, and W2). A least squares linear fit was applied to the data points (assuming linear trends) and the corresponding slope and intercept values obtained for these bands were: K$_{s}$=(-11.13$\pm$1.98, 81.29$\pm$7.47), W1=(-11.74$\pm$1.44, 90.08$\pm$5.45), and W2=(-13.35$\pm$3.30, 94.77$\pm$12.45), respectively. These trends show initial excess fractions of $\sim$80\% at the age of 1 Myr, decreasing down to about 20--30\% at the age of 6 Myrs. These results must be considered, however, under age uncertainties of $\sim$1--2 Myrs.

In addition, we wanted to compare inner disc dispersal evolution for the different types of disc classification (i.e. ``protoplanetary'', ``hybrid candidates'', and ``debris'' discs categories). For this we computed the fraction of stars per disc class for the three bands $<$10$\mu$m (without any age constraints). These are shown in Figure \ref{fig:discEvol}. We can see that 100\% of the ``protoplanetary'' discs have excess $<$10$\mu$m, as expected. About 84--89\% of the ``hybrid candidates'' discs have excess $<$10$\mu$m, and $\sim$26-34\% of the ``debris'' discs still have their inner excesses. This is further evidence that hybrid discs seem to be at an intermediate stage of evolution between protoplanetary and debris discs, with a small percentage of them having dissipated their inner region, while none of the protoplanetary discs show inner clearing and most of the debris discs do. This figure again shows similar excess fractions for the three wavelengths (2.2, 3.4 and 4.6$\mu$m), confirming the disc dispersal trends under 10$\mu$m we observe in Figure \ref{fig:fraction}.



\subsection{Comparison to low mass stars and other studies}
\label{sec:comparisonLMS}

In this section, we compare the results obtained for our sample of pre-main-sequence IMSs to previous studies of low mass stars (LMSs) and other studies of disc evolution. A key difference to the studies of LMSs is that such studies have the advantage of a well defined star forming region membership, and hence ratios of excess versus no-excess stars can be made, and also the pre-main-sequence nature of those stars suffers less ambiguity. \cite{Haisch2001}, \cite{Ribas2015} (from now on: R15), and \cite{Hernandez2010} investigated the disc evolution around solar and low mass stars and estimated disc dispersal timescales and the excess fraction dependence on stellar mass. Overall, these studies agree that the disc dispersal timescale has a certain dependence on stellar mass; discs seem to survive longer for lower mass stars, yet, most of them focus on the low-mass and solar mass regime. Therefore, we have attempted to perform a fair comparison of several studies against our IMSs sample. 

As a first comparison, we analysed three circumstellar disc surveys in star forming regions that have well characterized samples with available stellar parameters such as T$_{\rm eff}$, mass and age, and that have IR-excesses measured at 22$\mu$m in order to achieve a direct comparison to our sample. We took the sample of \cite{Andrews2013}: a study of disc dust mass vs. stellar mass dependence in the Taurus-Auriga star forming region, the sample of \cite{Barenfeld2016}: a study of LMSs with circumstellar discs in the Upper Scorpius OB Association, and the sample of \cite{Comeron2009}: a study of LMSs and their IR-excesses in the Lupus clouds from which we selected cloud 3 due to its larger number of targets. We collected the 2MASS and WISE photometric bands for these samples, and used their stellar parameters to select LMSs with masses $<$1.5M$_{\odot}$ and estimated their fractional excesses following the same procedure as we did in our study. For the sample of Taurus-Auriga, using the stellar age estimations of \cite{Andrews2013}, LMSs with ages $<$6 Myrs were selected. As done for our sample in Section \ref{sec:statistics}, the fraction of stars with excess at different wavelengths was then calculated for age bins of width 1 Myr. For the sample of Upper Scorpius, unfortunately, no individual ages were available, but this stellar association is in the age range 5--11 Myrs \citep{Barenfeld2016} and we included it for additional comparisons. For the sample of Lupus, although individual ages were available, most of them were aged between 1--3 Myrs and there were not enough stars to obtain excess fractions in the other age bins, therefore, we grouped them together in a single bin centered at 2 Myrs. For the three samples, we cross-matched the stars with Gaia DR3 distances and discarded any star having distances not compatible with that of the star forming region as inaccurate distances affect the measurement of the IR-excesses. This was particularly important for the sample of Lupus cloud 3 (at $\sim$200pc), where we found a considerable number of stars having distances in the order of 1000 to 10000 pc.

Figure \ref{fig:TauAur} shows the disc evolution for the LMSs in these three regions (analog to Figure \ref{fig:fraction} for our sample), this time considering stars of 0.08$M_{\odot}$ and 1.5$M_{\odot}$ as lower and upper limits for the distance range, with a T$_{\rm eff}$ range $\sim$2000K -- 7000K, and a R$_{\star}$ range $\sim$0.1 -- 1.7 R$_{\odot}$. We can see the difference in inner disc dispersal rates in LMSs with respect to IMSs. In Taurus-Auriga, the slopes for wavelengths $<$10$\mu$m range between $\sim$-6 to -4, while in our sample of IMSs, these slopes are much steeper, ranging between $\sim$-13 to -11. However, the most outstanding contrast is the difference in excess fractions among the three different wavelengths in LMSs, as opposed to the very similar fractions observed in IMSs. The excess fractions are consistently low at 2.2$\mu$m, and consistently high at 4.6$\mu$m, with the fraction at 3.4$\mu$m right in between. Similar fractions are observed in Upper Scorpius, where even at an older age there is a large difference between the fraction of excess at different wavelengths, at 2.2$\mu$m being very low, and at 4.6$\mu$m remaining very high. For the case of Lupus, the fractions are lower overall in comparison to Taurus-Auriga and Upper Scorpius, but the gap between the fraction of excess at different wavelengths is still observed. \cite{Comeron2009} comment on the puzzling properties of this population where there seems to be little circumstellar material within 1 au, and speculate that this might be due to formation "outside the shielding environment of a molecular cloud".  

Overall, our results show that the inner disc dispersal rate is slower in LMSs in comparison to IMSs and that the fraction of stars with excess is progressively lower closer to the star in LMSs. The lower initial fraction of excess at shorter wavelengths in LMSs is easy to explain if we consider a typical IMSs (for instance, a T$_{\rm eff}\sim$9700 and a R$_{\star}\sim$2.2 R$_{\odot}$); an excess at 2.2$\mu$m would mean the presence of dust at about 0.3 au from the star. But, for a LMSs (for example, T$_{\rm eff}\sim$3850 and a R$_{\star}\sim$0.6 R$_{\odot}$), excess at 2.2$\mu$m must come from dust at very few stellar radii or only $\sim$0.012 au from the star. The colder the star, the closest the dust must be in order to produce excess at a determined wavelength, then the coldest and lower mass stars are not expected to have excess at such short wavelengths. Radial drift is believed to fast dissipate the discs of most LMSs during the earlier stages \citep{Michel2021}, and thus the innermost regions are cleared already at a very early age.  

Theoretical studies such as \cite{Kunitomo2021} predict that discs around IMSs evolve faster than those around LMSs. Our results confirm that; the observed inner disc dispersal trends point towards IMSs dissipating their inner regions more rapidly than LMSs. This can be explained because of the fundamental differences between LMSs and IMSs on the pre-main sequence. Above 1.5 M$_{\odot}$, young
IMSs develop radiative envelopes, and consequently increase their luminosities and effective
temperatures (\citealt{PallaStahler1990}, \citealt{Siess2000}). This increases
far- and extreme-ultraviolet (FUV and EUV) emission, while loss of convection means that X-rays
are weaker \citep{WrightDrake2016}. This contributes to a rapid photoevaporative inner disc clearing process in IMSs, preventing the replenishment of material from the outer disc \citep{Wyatt2008}. On the other hand, for LMSs inner clearing seem to be mainly attributed to radial drift and rapid dust growth \citep{Michel2021}. 

As mentioned in the introduction, IMSs unlike LMSs, lack of low-mass close-in planets but have the highest frequency of giant planets (\citealt{Bowler2010}, \citealt{Reffert2015}). Giant exoplanets
are thought to form further away in the disc than the first few au \citep{Pollack1996}, and
then migrate inwards to their final orbits through the interaction with the disc (\citealt{Papaloizou2007}, \citealt{Paardekooper2010}, \citealt{YamadaInaba2011}). \cite{Kunitomo2011} showed that the missing planets could not have been engulfed by the host star as far out as 1 au, and their decreased frequency of detection must date back to the formation/migration stage. This is supported by previous studies of disc evolution such as \cite{KennedyKenyon2009} and R15: the IMSs disperse their discs faster than LMSs, and in this way halt inward-migrating giants at larger orbital distances. \cite{KennedyKenyon2009} studied disc dispersal dependence on stellar mass in several clusters by analysing disc fractions derived from the fraction of stars undergoing accretion. Unfortunately, this study is not directly comparable to our sample as all our stars have been pre-selected to have mid IR-excess. However, some comparisons can be made to the study of R15.

R15 compared disc evolution for a sample ranging from M to O spectral types (i.e. from low to high stellar masses). They split the sample in ``low-mass'' and ``high-mass'' at a limit of 2 M$_{\odot}$ and in age bins of 1-3 (``young'') and 3-11 Myrs (``old''). They defined three disc classification groups according to their IR-excesses in IRAC and MIPS1 photometry: protoplanetary (with excess at $\lambda\leq$ 10 $\mu$m, i.e. IRAC3 or IRAC4 bands), evolved (no excess in the IRAC bands but with excess in MIPS1), and discless (with no excess). We attempted to compare our sample with this study by performing the same disc classification and using the same definition of excess they proposed, which is based on an excess significance rather than a fractional excess. Since most of our sample does not have IRAC or MIPS1 observations, we used the closest WISE bands available instead (W2 instead of IRAC bands, and W4 in replacement of MIPS1). We divided our sample in the same mass and age bins as R15 and compared the resulting disc frequencies. However, since our sample only contains stars having some IR-excess level and since there are not many stars under 2 M$_{\odot}$ in our sample, we only considered the comparison in terms of the ratio between the protoplanetary and evolved discs (excluding the discless stars), and only in the ``high-mass'' bins (above 2 M$_{\odot}$). Despite the comparison not being able to test the stellar mass trends, our ratios are compatible with those of R15, observing similar percentages of young protoplanetary discs and confirming that the fraction of evolved discs increases with age.


\cite{Haisch2001} studied disc frequencies and lifetimes for several clusters of ages up to 30 Myrs and covering the entire stellar mass function. They used $JHKL$ colors to estimate circumstellar disc fractions for all the clusters. For the comparison, we used 2MASS $JHK_{s}$ photometry and WISE W1 band at 3.4 $\mu$m as it corresponds to the same wavelength as L band. JHKL color-color diagrams were used to derive the infrared excess fractions for our IMS sample following a procedure analogous to \cite{Haisch2001} but adapted to a bluer sample. Since the colder object in our sample corresponds to a spectral type F9, we chose the reddening limit to pass through the G0 color in \cite{Bessell_Brett1988}. \cite{Haisch2001} found that half of the stars lose their discs within $\sim$3 Myrs (with an initial disc fraction for the younger discs of about 80\%) and that overall all stars lose their discs within a timescale of about 6 Myrs. We found that, overall, the disc fractions were constant and close to a 100\% for all ages, and this is due to the selection criteria of our sample of having excess at 22$\mu$m. Our results based on the JHKL color-color diagram are then consistent with what we observed in our fractional excess analysis for wavelengths $>$ 10 $\mu$m. This suggests that the JHKL color-color methodology of IR-excess estimation might not be very sensitive to inner disc dissipation. 

Very similar results were obtained when comparing with the study by \cite{Hernandez2010}. They identified stars with excess at 24 $\mu$m as those with colours K - [24] > 0.69. For the comparison we used the closest bands K$_{s}$ and W4, and studied the excesses in our selection of pre-main-sequence IMSs. Again, we found that close to a 100\% of this sample had IR-excesses, which is consistent with the initial selection of the sample.

Finally, we searched our sample in the literature to find out how many of our objects have been previously studied and identified as stars with IR-excess. We crossmatched our sample with the census by \cite{Cotten2016} which is the most extensive compilation of infrared excess stars, with about 1750 sources including 246 references from the literature and new infrared excess stars, and also with more recent surveys of debris discs such as \cite{Pearce2022a}, and Herbig Ae/Be stars surveys such as e.g. \cite{Wichittanakom2020, Vioque2020, Vioque2022}. We found that 43 objects in our sample (34\%) have been previously identified as infrared excess stars. This means 66\% of our sample has no previous study of their IR-excess and that we likely identified 92 new young IMSs with measured mid-IR-excess. Those objects previously identified in the literature are marked with a `*' in Table \ref{tab:liter}.


\subsection{Discs caught between protoplanetary and debris disc stage}
\label{sec:hybrids}

\begin{figure}
	\includegraphics[width=\columnwidth]{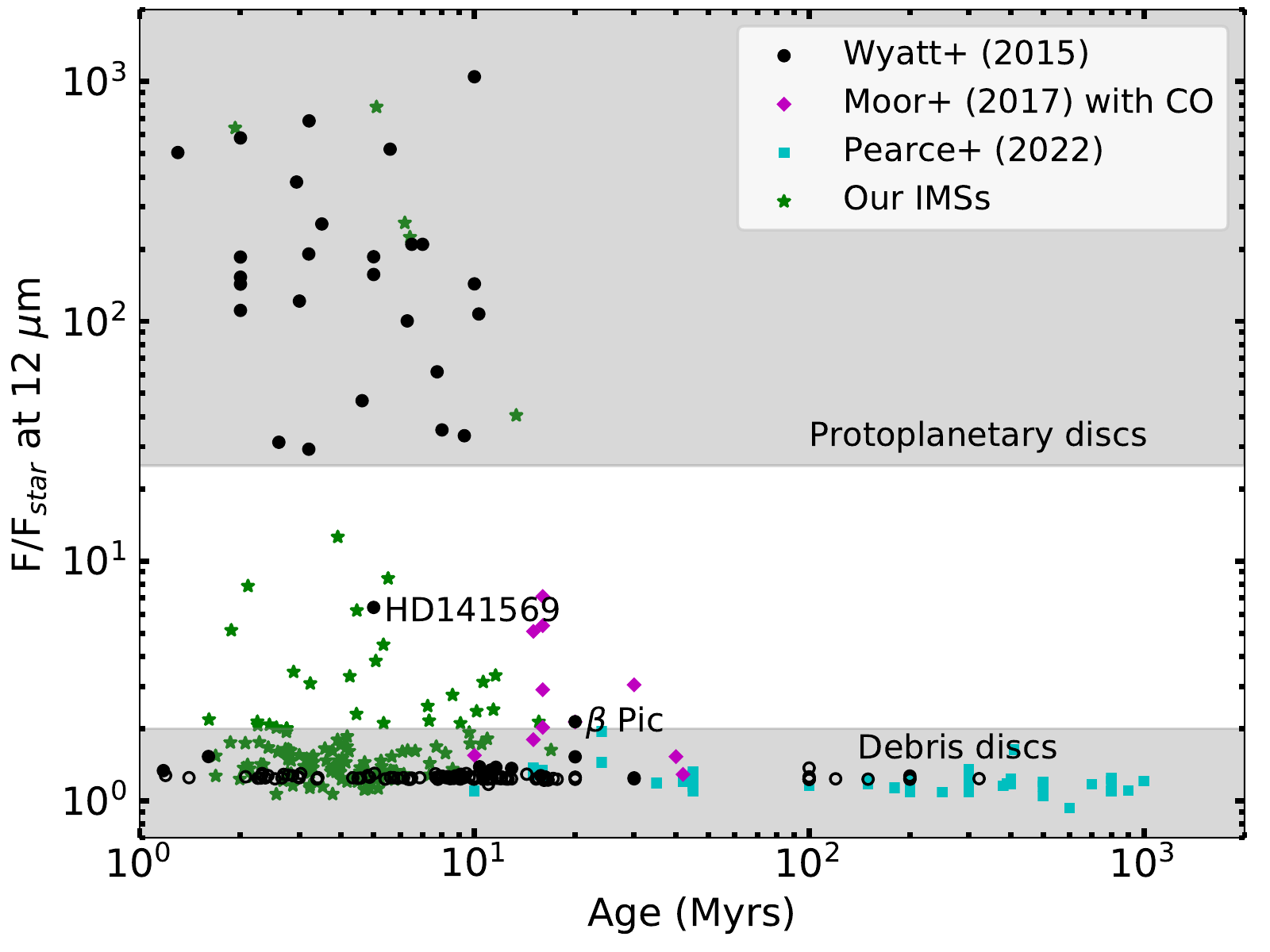}
    \caption{Comparison of $f_{12 \mu m}$ for our selection of IMSs with respect to previously known discs as shown in \protect\cite{Wyatt2015}, debris discs with CO from \protect\cite{Moor2017}, and studied debris discs from \protect\cite{Pearce2022a}. Filled circles are disc detections and open circles are upper limits. Gray shaded regions mark the excess level corresponding to protoplanetary and debris disc.}
    \label{fig:compWyatt}
\end{figure}

In Section \ref{sec:Classif} we classified our discs into `protoplanetary', `hybrid candidates', and `debris' categories based on their IR-excess level at 12 $\mu$m, $f_{12 \mu m}$. This is further illustrated in  Fig.  \ref{fig:compWyatt}  where we show our sample compared directly to the sample of W15, where it is evident how our newly identified pre-main sequence IMSs populate the previously nearly empty region between the debris and protoplanetary disc categories. The excess level and pre-main sequence nature of these objects suggest that they are akin to  HD\,141569; a well studied, peculiar object often classified as a hybrid disc due to its particular gas and dust properties, placing it in an intermediate state between protoplanetary and debris disc (\citealt{Miley2018}, \citealt{DiFolco2020}, \citealt{Gravity2021}). We find 17 IMSs that are `firmly' (above uncertainties) within the hybrid candidate range. 

This newly discovered population of hybrid disc candidates might be the key to unveil longly debated questions such as the origin of gas in debris discs around main sequence stars (15-50Myrs, \citealt{Moor2017}, \citealt{Pericaud2017}). This has been the subject of intense research over the past decade. An important clue to the origin is that gas in debris discs is detected almost exclusively around A-type stars, i.e. intermediate-mass stars (IMSs, M$_{\star}$=1.5-3.5M$_\odot$). This is why our `hybrid candidate' sample can be seen as the progenitors of the debris discs with gas detections during the main sequence. 
 
The primordial origin for the gas is one scenario, with remnant gas persisting from protoplanetary discs, i.e., from the pre-main sequence phase up to several tens of Myrs. The other scenario is the secondary origin, where the gas is produced via collisional erosion of planetesimals or sublimation of icy bodies such as comets, favoured scenario for debris discs with very tenuous amounts of gas, such as $\beta$ Pic \citep{Matra2018}. Such gas is shown to be able to enrich the atmosphere of planets with heavy elements, and influence their habitability, whereas primordial gas is not \citep{Kral2020}. The origin of gas in debris discs: primordial versus secondary, remains elusive (\citealt{Marino2020}, \citealt{Smirnov-Pinchukov2022}). The particular challenge for both scenarios is explaining the survival or sustained production of large masses of CO, comparable to low-mass protoplanetary discs, found at advanced ages above 10~Myr (\citealt{Kospal2013}, \citealt{Pericaud2017}, \citealt{Moor2017}).

A growing body of work has focused on the secondary origin for debris disc gas. However, a recent theoretical result of \cite{Nakatani2021} has shown that primordial origin of gas cannot be discarded in spite of the advanced age of gaseous debris discs (15-50Myr). In their simulations gas can survive well beyond the pre-main sequence, preferentially around IMSs. This happens when dust evolution in protoplanetary disc reaches the point where sub-micron sized dust is removed from the system, at the latest stage of disc photoevaporation. If this occurs while gas is still present, the central IMS, weak in X-rays, will be unable to heat the gas via FUV for which small dust is the conduit. In low-mass stars however, the gas is removed regardless, due to strong X-ray photoevaporation, which does not rely on dust as an intermediary. This hypothesis shifts the focus from collisional production of gas in debris discs to a much earlier stage, where the protoplanetary disc has just dispersed leaving behind the nascent debris disc system, such as in the case of our `hybrid candidate' sources.

To draw a comparison in terms of ages and excess levels to the debris discs with gas, in Fig. \ref{fig:compWyatt} we show the sample studied by \cite{Moor2017}. Using the stellar parameters published in that study (such as T$_{\rm eff}$ and age), we obtained their fractional excesses $f_{12 \mu m}$ and compared those debris discs around IMSs with CO detections to our sample, as shown in Figure \ref{fig:compWyatt}. We can see that most of these gaseous debris discs are found in the `hybrid candidate' category based on their $f_{12 \mu m}$ excesses, but, to the difference of our sample, theirs is older, with 15~Myr age for their youngest stars. This gives us a hint that our sample is ideally placed to investigate gas retention in discs around IMSs as proposed by \cite{Nakatani2021}, which will be the subject of our consequent observational study in a future publication. For further comparison, in Fig. \ref{fig:compWyatt} we also show the fractional excesses of known debris discs around IMSs with stellar parameters as published in \cite{Pearce2022a}. Their fractional excesses remain consistently below 2 across all ages.

\section{Summary and Conclusions}

We have studied and characterized the most homogeneous and unbiased sample of young intermediate mass stars with IR-excess to date with VLT/X-Shooter spectroscopic data, in the stellar mass range 1.5$\leq$M$_{\odot}\leq$3.5. Once identified, older contaminants likely evolved past the ZAMS were removed from the sample.

We have estimated all the basic stellar parameters for the full sample of candidates and studied their IR-excesses at 2.2, 3.4, 4.6, 12 and 22 $\rm{\mu}$m. We studied the age versus excess trends at these wavelengths. Since the initial selection criteria of the sample was having excess levels of W1-W4 $\geq$ 1, then, as expected, for the longer wavelengths ($>$ 10 $\mu $m) we found that close to a 100\% of the sources had fractional excesses $>1$, and in particular, a 100\% of them had excesses at 22 $\mu$m. 

Therefore, the key results regard wavelengths shorter than 10 $\rm{\mu}$m. At all three wavelengths, 2.2, 3.4 and 4.6~$\rm{\mu}$m, the IMSs show almost identical trends of decreasing excess fractions with age from about 80\% at 1--2~Myrs to $\sim$30\% at 5--6~Myrs. We also investigated these percentages for samples of known LMSs, pre-selecting these using the W1-W4 $\geq$ 1 excess criterion as in our sample. We chose a well characterized star forming region: Taurus-Auriga (sample from \citealt{Andrews2013}), collecting the stellar parameters from their catalogue and selecting the LMSs $<$ 1.5 M$_{\odot}$. For LMSs we find a very different behaviour, namely trends with much moderate slopes and excess fractions that differ for different wavelengths, with shortest wavelengths having consistently the lowest fractions. This would suggest that, while the inner disc regions with radius up to about 3 au dissipate  with age around IMSs, for LMSs the dissipation is more dependant on the proximity to the star. Perhaps the most marked difference is that most LMSs retain their 4.6$\rm{\mu}$m excess, which at 5-6~Myrs is still above 80\%. This means that as long as there is material in the outer disc (few to tens of au, mid-IR), disc material is retained also down at sub au-scale around LMSs. However, LMSs show larger degrees of dissipation (lower excess fractions) at 2.2$\rm{\mu}$m, where at any age the majority of stars in the sample show no excess. A caveat here is that we assume that Taurus is representative of LMSs. As a check we used Upper Scorpius (from \citealt{Barenfeld2016}) and Lupus cloud 3 (from \citealt{Comeron2009}). Upper Scorpius was similar in fractions to the oldest stars in Taurus at the age over 5 Myrs, confirming the trends observed in Taurus star forming region. The percentage of stars with excess were overall a bit lower for the case of Lupus, but the drop in excess fractions for shorter wavelengths (i.e. closer to the star) was confirmed.

This empirical study confirms that circumstellar discs around IMSs evolve differently than those surrounding LMSs and this is likely the cause of the distinct planetary systems architectures observed in IMSs as compared to their low mass counterparts. The faster inner disc dispersal in IMSs might explain the lack of low mass short period planets around these stars, a phenomenon that is not observed in LMSs.
 
 Based on the fractional excesses at 12 $\rm{\mu}$m, we classified the sample into `protoplanetary' (5 objects), `hybrid candidates' (17 objects), and `debris' (112 objects) disc categories. We report here for the first time a new population of 17 hybrid disc candidates that might represent a missing link in the evolution from protoplanetary to debris discs. These stars are the most likely progenitors of the known debris discs with gas at the main sequence and as such are well placed to test the primordial gas hypothesis for the origin of gas in those discs. 
 
 Several Herbig Ae/Be stars undergoing accretion were found in the preliminary sample and we estimated their mass accretion rates and compared it to LMSs studies, finding agreement with the mass vs. mass accretion rate correlation in LMSs and overall trends found in other studies (please see the Appendix \ref{sec:accretion} for details).
 
In addition, one of the most remarkable findings in this paper is that we identified a dispersed population of young IMSs, the majority not associated with any previously known region of star formation and not identified in any previous surveys of IR-excess stars. This implies the possibility that IMSs can form isolated. 

We will further analyse our sample of IMSs, addressing other factors playing a role in disc evolution such as binarity and the presence of gas.
We will study the hybrid disc candidates with ALMA\footnote{Project 2022.1.01686.S, PI: Pani\'c} and high-resolution spectroscopic observations\footnote{ESO UVES Programmes 109.23K8.001 and 110.23YZ.001, PI: Iglesias}, aiming to characterize these discs, their morphology, binarity, composition, and presence of gas.

\section*{Acknowledgements}

D.P.I. and O.P. acknowledge support from the Science and Technology Facilities Council via grant number ST/T000287/1. LS is senior FNRS researcher. The authors thank Grant Kennedy for helpful discussions. The authors thank the anonymous referee for the useful feedback that helped improve the paper.
This publication makes use of data products from the Wide-field Infrared Survey Explorer, which is a joint project of the University of California, Los Angeles, and the Jet Propulsion Laboratory/California Institute of Technology, funded by the National Aeronautics and Space Administration. This publication makes use of data products from the Two Micron All Sky Survey, which is a joint project of the University of Massachusetts and the Infrared Processing and Analysis Center/California Institute of Technology, funded by the National Aeronautics and Space Administration and the National Science Foundation. This work has made use of data from the European Space Agency (ESA) mission
{\it Gaia} (\url{https://www.cosmos.esa.int/gaia}), processed by the {\it Gaia}
Data Processing and Analysis Consortium (DPAC,
\url{https://www.cosmos.esa.int/web/gaia/dpac/consortium}). Funding for the DPAC
has been provided by national institutions, in particular the institutions
participating in the {\it Gaia} Multilateral Agreement. Data from the following ESO X-Shooter programmes have been used in this work: 0101.C-0866(A), 0101.C-0902(A), 0102.C-0882(A), 0103.C-0887(B), 084.C-0952(A), 085.C-0764(A), 088.B-0485(A), 088.C-0218(A), 088.C-0218(B), 088.C-0218(C), 088.C-0218(E), 089.C-0874(A), 091.D-0905(A), 093.D-0415(A), 093.D-0415(B), 097.C-0378(A), 189.B-0925(A), 385.C-0131(A), 60.A-9022(C).

\section*{Data Availability}

The data underlying this article are available in the ESO archive at \url{http://archive.eso.org/cms.html}, in the Gaia archive at \url{https://gea.esac.esa.int/archive/}, in the 2MASS catalog at \url{https://irsa.ipac.caltech.edu/Missions/2mass.html}, in the allWISE catalog at \url{https://wise2.ipac.caltech.edu/docs/release/allwise/}, and PARSEC stellar tracks and isochrones are available at \url{https://people.sissa.it/~sbressan/parsec.html}. The datasets were derived from sources in the public domain: Tycho-2 catalogue at \url{https://www.cosmos.esa.int/web/hipparcos/tycho-2}, Hipparcos catalogue at \url{https://cdsarc.cds.unistra.fr/viz-bin/cat/I/311}, and TGAS catalogue at \url{https://cdsarc.cds.unistra.fr/viz-bin/cat/I/337}.



\bibliographystyle{mnras}
\bibliography{biblio} 

\begin{thebibliography}{}
\makeatletter
\relax
\def\mn@urlcharsother{\let\do\@makeother \do\$\do\&\do\#\do\^\do\_\do\%\do\~}
\def\mn@doi{\begingroup\mn@urlcharsother \@ifnextchar [ {\mn@doi@}
  {\mn@doi@[]}}
\def\mn@doi@[#1]#2{\def\@tempa{#1}\ifx\@tempa\@empty \href
  {http://dx.doi.org/#2} {doi:#2}\else \href {http://dx.doi.org/#2} {#1}\fi
  \endgroup}
\def\mn@eprint#1#2{\mn@eprint@#1:#2::\@nil}
\def\mn@eprint@arXiv#1{\href {http://arxiv.org/abs/#1} {{\tt arXiv:#1}}}
\def\mn@eprint@dblp#1{\href {http://dblp.uni-trier.de/rec/bibtex/#1.xml}
  {dblp:#1}}
\def\mn@eprint@#1:#2:#3:#4\@nil{\def\@tempa {#1}\def\@tempb {#2}\def\@tempc
  {#3}\ifx \@tempc \@empty \let \@tempc \@tempb \let \@tempb \@tempa \fi \ifx
  \@tempb \@empty \def\@tempb {arXiv}\fi \@ifundefined
  {mn@eprint@\@tempb}{\@tempb:\@tempc}{\expandafter \expandafter \csname
  mn@eprint@\@tempb\endcsname \expandafter{\@tempc}}}

\bibitem[\protect\citeauthoryear{{Andrews}, {Rosenfeld}, {Kraus}  \&
  {Wilner}}{{Andrews} et~al.}{2013}]{Andrews2013}
{Andrews} S.~M.,  {Rosenfeld} K.~A.,  {Kraus} A.~L.,   {Wilner} D.~J.,  2013,
  \mn@doi [\apj] {10.1088/0004-637X/771/2/129}, \href
  {https://ui.adsabs.harvard.edu/abs/2013ApJ...771..129A} {771, 129}

\bibitem[\protect\citeauthoryear{{Arun}, {Mathew}, {Manoj}, {Ujjwal}, {Kartha},
  {Viswanath}, {Narang}  \& {Paul}}{{Arun} et~al.}{2019}]{Arun2019}
{Arun} R.,  {Mathew} B.,  {Manoj} P.,  {Ujjwal} K.,  {Kartha} S.~S.,
  {Viswanath} G.,  {Narang} M.,   {Paul} K.~T.,  2019, \mn@doi [\aj]
  {10.3847/1538-3881/ab0ca1}, \href
  {https://ui.adsabs.harvard.edu/abs/2019AJ....157..159A} {157, 159}

\bibitem[\protect\citeauthoryear{{Bailer-Jones}, {Rybizki}, {Fouesneau},
  {Demleitner}  \& {Andrae}}{{Bailer-Jones} et~al.}{2021}]{Bailer-Jones2021}
{Bailer-Jones} C.~A.~L.,  {Rybizki} J.,  {Fouesneau} M.,  {Demleitner} M.,
  {Andrae} R.,  2021, \mn@doi [\aj] {10.3847/1538-3881/abd806}, \href
  {https://ui.adsabs.harvard.edu/abs/2021AJ....161..147B} {161, 147}

\bibitem[\protect\citeauthoryear{{Baranne}, {Mayor}  \& {Poncet}}{{Baranne}
  et~al.}{1979}]{Baranne1979}
{Baranne} A.,  {Mayor} M.,   {Poncet} J.~L.,  1979, \mn@doi [Vistas in
  Astronomy] {10.1016/0083-6656(79)90016-3}, \href
  {https://ui.adsabs.harvard.edu/abs/1979VA.....23..279B} {23, 279}

\bibitem[\protect\citeauthoryear{{Barenfeld}, {Carpenter}, {Ricci}  \&
  {Isella}}{{Barenfeld} et~al.}{2016}]{Barenfeld2016}
{Barenfeld} S.~A.,  {Carpenter} J.~M.,  {Ricci} L.,   {Isella} A.,  2016,
  \mn@doi [\apj] {10.3847/0004-637X/827/2/142}, \href
  {https://ui.adsabs.harvard.edu/abs/2016ApJ...827..142B} {827, 142}

\bibitem[\protect\citeauthoryear{{Bell}, {Mamajek}  \& {Naylor}}{{Bell}
  et~al.}{2015}]{Bell2015}
{Bell} C. P.~M.,  {Mamajek} E.~E.,   {Naylor} T.,  2015, \mn@doi [\mnras]
  {10.1093/mnras/stv1981}, \href
  {https://ui.adsabs.harvard.edu/abs/2015MNRAS.454..593B} {454, 593}

\bibitem[\protect\citeauthoryear{{Bessell} \& {Brett}}{{Bessell} \&
  {Brett}}{1988}]{Bessell_Brett1988}
{Bessell} M.~S.,  {Brett} J.~M.,  1988, \mn@doi [\pasp] {10.1086/132281}, \href
  {https://ui.adsabs.harvard.edu/abs/1988PASP..100.1134B} {100, 1134}

\bibitem[\protect\citeauthoryear{{Bessell}, {Castelli}  \& {Plez}}{{Bessell}
  et~al.}{1998}]{Bessell1998}
{Bessell} M.~S.,  {Castelli} F.,   {Plez} B.,  1998, \aap, \href
  {https://ui.adsabs.harvard.edu/abs/1998A&A...333..231B} {333, 231}

\bibitem[\protect\citeauthoryear{{Bowler} et~al.,}{{Bowler}
  et~al.}{2010}]{Bowler2010}
{Bowler} B.~P.,  et~al., 2010, \mn@doi [\apj] {10.1088/0004-637X/709/1/396},
  \href {https://ui.adsabs.harvard.edu/abs/2010ApJ...709..396B} {709, 396}

\bibitem[\protect\citeauthoryear{{Bressan}, {Marigo}, {Girardi}, {Salasnich},
  {Dal Cero}, {Rubele}  \& {Nanni}}{{Bressan} et~al.}{2012}]{Bressan2012}
{Bressan} A.,  {Marigo} P.,  {Girardi} L.,  {Salasnich} B.,  {Dal Cero} C.,
  {Rubele} S.,   {Nanni} A.,  2012, \mn@doi [\mnras]
  {10.1111/j.1365-2966.2012.21948.x}, \href
  {https://ui.adsabs.harvard.edu/abs/2012MNRAS.427..127B} {427, 127}

\bibitem[\protect\citeauthoryear{{Cardelli}, {Clayton}  \& {Mathis}}{{Cardelli}
  et~al.}{1989}]{Cardelli1989}
{Cardelli} J.~A.,  {Clayton} G.~C.,   {Mathis} J.~S.,  1989, \mn@doi [\apj]
  {10.1086/167900}, \href
  {https://ui.adsabs.harvard.edu/abs/1989ApJ...345..245C} {345, 245}

\bibitem[\protect\citeauthoryear{{Castelli}, {Gratton}  \& {Kurucz}}{{Castelli}
  et~al.}{1997}]{Castelli1997}
{Castelli} F.,  {Gratton} R.~G.,   {Kurucz} R.~L.,  1997, \aap, \href
  {http://adsabs.harvard.edu/abs/1997A%26A...318..841C} {318, 841}

\bibitem[\protect\citeauthoryear{{Chauvin} et~al.,}{{Chauvin}
  et~al.}{2018}]{Chauvin2018}
{Chauvin} G.,  et~al., 2018, \mn@doi [\aap] {10.1051/0004-6361/201732077},
  \href {https://ui.adsabs.harvard.edu/abs/2018A&A...617A..76C} {617, A76}

\bibitem[\protect\citeauthoryear{{Chen}, {Shan}  \& {Zhang}}{{Chen}
  et~al.}{2016}]{Chen2016}
{Chen} P.~S.,  {Shan} H.~G.,   {Zhang} P.,  2016, \mn@doi [\na]
  {10.1016/j.newast.2015.09.001}, \href
  {https://ui.adsabs.harvard.edu/abs/2016NewA...44....1C} {44, 1}

\bibitem[\protect\citeauthoryear{{Comer{\'o}n}, {Spezzi}  \& {L{\'o}pez
  Mart{\'\i}}}{{Comer{\'o}n} et~al.}{2009}]{Comeron2009}
{Comer{\'o}n} F.,  {Spezzi} L.,   {L{\'o}pez Mart{\'\i}} B.,  2009, \mn@doi
  [\aap] {10.1051/0004-6361/200911771}, \href
  {https://ui.adsabs.harvard.edu/abs/2009A&A...500.1045C} {500, 1045}

\bibitem[\protect\citeauthoryear{{Cotten} \& {Song}}{{Cotten} \&
  {Song}}{2016}]{Cotten2016}
{Cotten} T.~H.,  {Song} I.,  2016, \mn@doi [\apjs]
  {10.3847/0067-0049/225/1/15}, \href
  {https://ui.adsabs.harvard.edu/abs/2016ApJS..225...15C} {225, 15}

\bibitem[\protect\citeauthoryear{{Cutri} et~al.,}{{Cutri}
  et~al.}{2021}]{Cutri2014}
{Cutri} R.~M.,  et~al., 2021, VizieR Online Data Catalog, \href
  {https://ui.adsabs.harvard.edu/abs/2014yCat.2328....0C} {p. II/328}

\bibitem[\protect\citeauthoryear{{Dahm}}{{Dahm}}{2015}]{Dahm2015}
{Dahm} S.~E.,  2015, \mn@doi [\apj] {10.1088/0004-637X/813/2/108}, \href
  {https://ui.adsabs.harvard.edu/abs/2015ApJ...813..108D} {813, 108}

\bibitem[\protect\citeauthoryear{{Damiani} et~al.,}{{Damiani}
  et~al.}{2014}]{Damiani2014}
{Damiani} F.,  et~al., 2014, \mn@doi [\aap] {10.1051/0004-6361/201323306},
  \href {https://ui.adsabs.harvard.edu/abs/2014A&A...566A..50D} {566, A50}

\bibitem[\protect\citeauthoryear{{Di Folco}, {P{\'e}ricaud}, {Dutrey},
  {Augereau}, {Chapillon}, {Guilloteau}, {Pi{\'e}tu}  \& {Boccaletti}}{{Di
  Folco} et~al.}{2020}]{DiFolco2020}
{Di Folco} E.,  {P{\'e}ricaud} J.,  {Dutrey} A.,  {Augereau} J.~C.,
  {Chapillon} E.,  {Guilloteau} S.,  {Pi{\'e}tu} V.,   {Boccaletti} A.,  2020,
  \mn@doi [\aap] {10.1051/0004-6361/201732243}, \href
  {https://ui.adsabs.harvard.edu/abs/2020A&A...635A..94D} {635, A94}

\bibitem[\protect\citeauthoryear{{Dobbie}, {Lodieu}  \& {Sharp}}{{Dobbie}
  et~al.}{2010}]{Dobbie2010}
{Dobbie} P.~D.,  {Lodieu} N.,   {Sharp} R.~G.,  2010, \mn@doi [\mnras]
  {10.1111/j.1365-2966.2010.17355.x}, \href
  {https://ui.adsabs.harvard.edu/abs/2010MNRAS.409.1002D} {409, 1002}

\bibitem[\protect\citeauthoryear{{Elliott}, {Bayo}, {Melo}, {Torres}, {Sterzik}
   \& {Quast}}{{Elliott} et~al.}{2014}]{Elliott2014}
{Elliott} P.,  {Bayo} A.,  {Melo} C.~H.~F.,  {Torres} C.~A.~O.,  {Sterzik} M.,
   {Quast} G.~R.,  2014, \mn@doi [\aap] {10.1051/0004-6361/201423856}, \href
  {https://ui.adsabs.harvard.edu/abs/2014A&A...568A..26E} {568, A26}

\bibitem[\protect\citeauthoryear{{Ercolano}, {Bastian}, {Spezzi}  \&
  {Owen}}{{Ercolano} et~al.}{2011}]{Ercolano2011}
{Ercolano} B.,  {Bastian} N.,  {Spezzi} L.,   {Owen} J.,  2011, \mn@doi
  [\mnras] {10.1111/j.1365-2966.2011.19051.x}, \href
  {https://ui.adsabs.harvard.edu/abs/2011MNRAS.416..439E} {416, 439}

\bibitem[\protect\citeauthoryear{{Fabregat} \& {Torrej{\'o}n}}{{Fabregat} \&
  {Torrej{\'o}n}}{2000}]{Fabregat_Torrejon2000}
{Fabregat} J.,  {Torrej{\'o}n} J.~M.,  2000, \aap, \href
  {https://ui.adsabs.harvard.edu/abs/2000A&A...357..451F} {357, 451}

\bibitem[\protect\citeauthoryear{{Fairlamb}, {Oudmaijer}, {Mendigut{\'\i}a},
  {Ilee}  \& {van den Ancker}}{{Fairlamb} et~al.}{2015}]{Fairlamb2015}
{Fairlamb} J.~R.,  {Oudmaijer} R.~D.,  {Mendigut{\'\i}a} I.,  {Ilee} J.~D.,
  {van den Ancker} M.~E.,  2015, \mn@doi [\mnras] {10.1093/mnras/stv1576},
  \href {https://ui.adsabs.harvard.edu/abs/2015MNRAS.453..976F} {453, 976}

\bibitem[\protect\citeauthoryear{{Fairlamb}, {Oudmaijer}, {Mendigutia}, {Ilee}
  \& {van den Ancker}}{{Fairlamb} et~al.}{2017}]{Fairlamb2017}
{Fairlamb} J.~R.,  {Oudmaijer} R.~D.,  {Mendigutia} I.,  {Ilee} J.~D.,   {van
  den Ancker} M.~E.,  2017, \mn@doi [\mnras] {10.1093/mnras/stw2643}, \href
  {https://ui.adsabs.harvard.edu/abs/2017MNRAS.464.4721F} {464, 4721}

\bibitem[\protect\citeauthoryear{{Frasca}, {Biazzo}, {Alcal{\'a}}, {Manara},
  {Stelzer}, {Covino}  \& {Antoniucci}}{{Frasca} et~al.}{2017}]{Frasca2017}
{Frasca} A.,  {Biazzo} K.,  {Alcal{\'a}} J.~M.,  {Manara} C.~F.,  {Stelzer} B.,
   {Covino} E.,   {Antoniucci} S.,  2017, \mn@doi [\aap]
  {10.1051/0004-6361/201630108}, \href
  {https://ui.adsabs.harvard.edu/abs/2017A&A...602A..33F} {602, A33}

\bibitem[\protect\citeauthoryear{{Freudling}, {Romaniello}, {Bramich},
  {Ballester}, {Forchi}, {Garc{\'{\i}}a-Dabl{\'o}}, {Moehler}  \&
  {Neeser}}{{Freudling} et~al.}{2013}]{Freudling2013}
{Freudling} W.,  {Romaniello} M.,  {Bramich} D.~M.,  {Ballester} P.,  {Forchi}
  V.,  {Garc{\'{\i}}a-Dabl{\'o}} C.~E.,  {Moehler} S.,   {Neeser} M.~J.,  2013,
  \mn@doi [\aap] {10.1051/0004-6361/201322494}, \href
  {http://adsabs.harvard.edu/abs/2013A%26A...559A..96F} {559, A96}

\bibitem[\protect\citeauthoryear{{Gagn{\'e}} et~al.,}{{Gagn{\'e}}
  et~al.}{2018}]{Gagne2018}
{Gagn{\'e}} J.,  et~al., 2018, \mn@doi [\apj] {10.3847/1538-4357/aaae09}, \href
  {https://ui.adsabs.harvard.edu/abs/2018ApJ...856...23G} {856, 23}

\bibitem[\protect\citeauthoryear{{Gagn{\'e}}, {David}, {Mamajek}, {Mann},
  {Faherty}  \& {B{\'e}dard}}{{Gagn{\'e}} et~al.}{2020}]{Gagne2020}
{Gagn{\'e}} J.,  {David} T.~J.,  {Mamajek} E.~E.,  {Mann} A.~W.,  {Faherty}
  J.~K.,   {B{\'e}dard} A.,  2020, \mn@doi [\apj] {10.3847/1538-4357/abb77e},
  \href {https://ui.adsabs.harvard.edu/abs/2020ApJ...903...96G} {903, 96}

\bibitem[\protect\citeauthoryear{{Gaia Collaboration}}{{Gaia
  Collaboration}}{2018}]{GaiaDR2_2018}
{Gaia Collaboration} 2018, VizieR Online Data Catalog, \href
  {https://ui.adsabs.harvard.edu/abs/2018yCat.1345....0G} {p. I/345}

\bibitem[\protect\citeauthoryear{{Gaia Collaboration} et~al.,}{{Gaia
  Collaboration} et~al.}{2016}]{Gaia2016}
{Gaia Collaboration} et~al., 2016, \mn@doi [\aap]
  {10.1051/0004-6361/201629512}, \href
  {http://adsabs.harvard.edu/abs/2016A%26A...595A...2G} {595, A2}

\bibitem[\protect\citeauthoryear{{Gaia Collaboration} et~al.,}{{Gaia
  Collaboration} et~al.}{2022}]{Gaia2022k}
{Gaia Collaboration} et~al., 2022, arXiv e-prints, \href
  {https://ui.adsabs.harvard.edu/abs/2022arXiv220800211G} {p. arXiv:2208.00211}

\bibitem[\protect\citeauthoryear{{Gaudi}}{{Gaudi}}{2022}]{Gaudi2022}
{Gaudi} B.~S.,  2022, in {Biazzo} K.,  {Bozza} V.,  {Mancini} L.,   {Sozzetti}
  A.,  eds,  Astrophysics and Space Science Library Vol. 466, Demographics of
  Exoplanetary Systems, Lecture Notes of the 3rd Advanced School on
  Exoplanetary Science. pp 237--291 (\mn@eprint {arXiv} {2102.01715}),
  \mn@doi{10.1007/978-3-030-88124-5\_4}

\bibitem[\protect\citeauthoryear{{Gravity Collaboration} et~al.,}{{Gravity
  Collaboration} et~al.}{2021}]{Gravity2021}
{Gravity Collaboration} et~al., 2021, \mn@doi [\aap]
  {10.1051/0004-6361/202141103}, \href
  {https://ui.adsabs.harvard.edu/abs/2021A&A...655A.112G} {655, A112}

\bibitem[\protect\citeauthoryear{{Gray} \& {Corbally}}{{Gray} \&
  {Corbally}}{2009}]{GrayCorbally2009}
{Gray} R.~O.,  {Corbally} J. C.,  2009, {Stellar Spectral Classification}.
Princeton University Press

\bibitem[\protect\citeauthoryear{{Guzm{\'a}n-D{\'\i}az}
  et~al.,}{{Guzm{\'a}n-D{\'\i}az} et~al.}{2021}]{GuzmanDiaz2021}
{Guzm{\'a}n-D{\'\i}az} J.,  et~al., 2021, \mn@doi [\aap]
  {10.1051/0004-6361/202039519}, \href
  {https://ui.adsabs.harvard.edu/abs/2021A&A...650A.182G} {650, A182}

\bibitem[\protect\citeauthoryear{{Haisch}, {Lada}  \& {Lada}}{{Haisch}
  et~al.}{2001}]{Haisch2001}
{Haisch} Karl~E. J.,  {Lada} E.~A.,   {Lada} C.~J.,  2001, \mn@doi [\apjl]
  {10.1086/320685}, \href
  {https://ui.adsabs.harvard.edu/abs/2001ApJ...553L.153H} {553, L153}

\bibitem[\protect\citeauthoryear{{Hern{\'a}ndez}, {Calvet}, {Brice{\~n}o},
  {Hartmann}  \& {Berlind}}{{Hern{\'a}ndez} et~al.}{2004}]{Hernandez2004}
{Hern{\'a}ndez} J.,  {Calvet} N.,  {Brice{\~n}o} C.,  {Hartmann} L.,
  {Berlind} P.,  2004, \mn@doi [\aj] {10.1086/381908}, \href
  {https://ui.adsabs.harvard.edu/abs/2004AJ....127.1682H} {127, 1682}

\bibitem[\protect\citeauthoryear{{Hern{\'a}ndez}, {Morales-Calderon}, {Calvet},
  {Hartmann}, {Muzerolle}, {Gutermuth}, {Luhman}  \&
  {Stauffer}}{{Hern{\'a}ndez} et~al.}{2010}]{Hernandez2010}
{Hern{\'a}ndez} J.,  {Morales-Calderon} M.,  {Calvet} N.,  {Hartmann} L.,
  {Muzerolle} J.,  {Gutermuth} R.,  {Luhman} K.~L.,   {Stauffer} J.,  2010,
  \mn@doi [\apj] {10.1088/0004-637X/722/2/1226}, \href
  {https://ui.adsabs.harvard.edu/abs/2010ApJ...722.1226H} {722, 1226}

\bibitem[\protect\citeauthoryear{{Hillenbrand}, {Strom}, {Vrba}  \&
  {Keene}}{{Hillenbrand} et~al.}{1992}]{Hillenbrand1992}
{Hillenbrand} L.~A.,  {Strom} S.~E.,  {Vrba} F.~J.,   {Keene} J.,  1992,
  \mn@doi [\apj] {10.1086/171819}, \href
  {https://ui.adsabs.harvard.edu/abs/1992ApJ...397..613H} {397, 613}

\bibitem[\protect\citeauthoryear{{Iglesias} et~al.,}{{Iglesias}
  et~al.}{2018}]{Iglesias2018}
{Iglesias} D.,  et~al., 2018, \mn@doi [\mnras] {10.1093/mnras/sty1724}, \href
  {https://ui.adsabs.harvard.edu/abs/2018MNRAS.480..488I} {480, 488}

\bibitem[\protect\citeauthoryear{{Johnson} et~al.,}{{Johnson}
  et~al.}{2007}]{Johnson2007b}
{Johnson} J.~A.,  et~al., 2007, \mn@doi [\apj] {10.1086/519677}, \href
  {https://ui.adsabs.harvard.edu/abs/2007ApJ...665..785J} {665, 785}

\bibitem[\protect\citeauthoryear{{Johnson}, {Howard}, {Bowler}, {Henry},
  {Marcy}, {Wright}, {Fischer}  \& {Isaacson}}{{Johnson}
  et~al.}{2010}]{Johnson2010}
{Johnson} J.~A.,  {Howard} A.~W.,  {Bowler} B.~P.,  {Henry} G.~W.,  {Marcy}
  G.~W.,  {Wright} J.~T.,  {Fischer} D.~A.,   {Isaacson} H.,  2010, \mn@doi
  [\pasp] {10.1086/653809}, \href
  {https://ui.adsabs.harvard.edu/abs/2010PASP..122..701J} {122, 701}

\bibitem[\protect\citeauthoryear{{Kausch} et~al.,}{{Kausch}
  et~al.}{2015}]{Kausch2015}
{Kausch} W.,  et~al., 2015, \mn@doi [\aap] {10.1051/0004-6361/201423909}, \href
  {http://adsabs.harvard.edu/abs/2015A%26A...576A..78K} {576, A78}

\bibitem[\protect\citeauthoryear{{Kennedy} \& {Kenyon}}{{Kennedy} \&
  {Kenyon}}{2009}]{KennedyKenyon2009}
{Kennedy} G.~M.,  {Kenyon} S.~J.,  2009, \mn@doi [\apj]
  {10.1088/0004-637X/695/2/1210}, \href
  {https://ui.adsabs.harvard.edu/abs/2009ApJ...695.1210K} {695, 1210}

\bibitem[\protect\citeauthoryear{{K{\'o}sp{\'a}l} et~al.,}{{K{\'o}sp{\'a}l}
  et~al.}{2013}]{Kospal2013}
{K{\'o}sp{\'a}l} {\'A}.,  et~al., 2013, \mn@doi [\apj]
  {10.1088/0004-637X/776/2/77}, \href
  {http://adsabs.harvard.edu/abs/2013ApJ...776...77K} {776, 77}

\bibitem[\protect\citeauthoryear{{Kral}, {Davoult}  \& {Charnay}}{{Kral}
  et~al.}{2020}]{Kral2020}
{Kral} Q.,  {Davoult} J.,   {Charnay} B.,  2020, \mn@doi [Nature Astronomy]
  {10.1038/s41550-020-1050-2}, \href
  {https://ui.adsabs.harvard.edu/abs/2020NatAs...4..769K} {4, 769}

\bibitem[\protect\citeauthoryear{{Kretke}, {Lin}, {Garaud}  \&
  {Turner}}{{Kretke} et~al.}{2009}]{Kretke2009}
{Kretke} K.~A.,  {Lin} D.~N.~C.,  {Garaud} P.,   {Turner} N.~J.,  2009, \mn@doi
  [\apj] {10.1088/0004-637X/690/1/407}, \href
  {https://ui.adsabs.harvard.edu/abs/2009ApJ...690..407K} {690, 407}

\bibitem[\protect\citeauthoryear{{Kunitomo}, {Ikoma}, {Sato}, {Katsuta}  \&
  {Ida}}{{Kunitomo} et~al.}{2011}]{Kunitomo2011}
{Kunitomo} M.,  {Ikoma} M.,  {Sato} B.,  {Katsuta} Y.,   {Ida} S.,  2011,
  \mn@doi [\apj] {10.1088/0004-637X/737/2/66}, \href
  {https://ui.adsabs.harvard.edu/abs/2011ApJ...737...66K} {737, 66}

\bibitem[\protect\citeauthoryear{{Kunitomo}, {Ida}, {Takeuchi}, {Pani{\'c}},
  {Miley}  \& {Suzuki}}{{Kunitomo} et~al.}{2021}]{Kunitomo2021}
{Kunitomo} M.,  {Ida} S.,  {Takeuchi} T.,  {Pani{\'c}} O.,  {Miley} J.~M.,
  {Suzuki} T.~K.,  2021, \mn@doi [\apj] {10.3847/1538-4357/abdb2a}, \href
  {https://ui.adsabs.harvard.edu/abs/2021ApJ...909..109K} {909, 109}

\bibitem[\protect\citeauthoryear{{Kurtz} \& {Mink}}{{Kurtz} \&
  {Mink}}{1998}]{Kurtz1998}
{Kurtz} M.~J.,  {Mink} D.~J.,  1998, \mn@doi [\pasp] {10.1086/316207}, \href
  {https://ui.adsabs.harvard.edu/abs/1998PASP..110..934K} {110, 934}

\bibitem[\protect\citeauthoryear{{Lagrange} et~al.,}{{Lagrange}
  et~al.}{2019}]{Lagrange2019}
{Lagrange} A.~M.,  et~al., 2019, \mn@doi [Nature Astronomy]
  {10.1038/s41550-019-0857-1}, \href
  {https://ui.adsabs.harvard.edu/abs/2019NatAs...3.1135L} {3, 1135}

\bibitem[\protect\citeauthoryear{{Linsky}}{{Linsky}}{2017}]{Linsky2017}
{Linsky} J.~L.,  2017, \mn@doi [\araa] {10.1146/annurev-astro-091916-055327},
  \href {https://ui.adsabs.harvard.edu/abs/2017ARA&A..55..159L} {55, 159}

\bibitem[\protect\citeauthoryear{{Lovis} \& {Mayor}}{{Lovis} \&
  {Mayor}}{2007}]{Lovis_Mayor2007}
{Lovis} C.,  {Mayor} M.,  2007, \mn@doi [\aap] {10.1051/0004-6361:20077375},
  \href {https://ui.adsabs.harvard.edu/abs/2007A&A...472..657L} {472, 657}

\bibitem[\protect\citeauthoryear{{Manara} et~al.,}{{Manara}
  et~al.}{2017}]{Manara2017}
{Manara} C.~F.,  et~al., 2017, \mn@doi [\aap] {10.1051/0004-6361/201630147},
  \href {https://ui.adsabs.harvard.edu/abs/2017A&A...604A.127M} {604, A127}

\bibitem[\protect\citeauthoryear{{Marino}, {Flock}, {Henning}, {Kral},
  {Matr{\`a}}  \& {Wyatt}}{{Marino} et~al.}{2020}]{Marino2020}
{Marino} S.,  {Flock} M.,  {Henning} T.,  {Kral} Q.,  {Matr{\`a}} L.,   {Wyatt}
  M.~C.,  2020, \mn@doi [\mnras] {10.1093/mnras/stz3487}, \href
  {https://ui.adsabs.harvard.edu/abs/2020MNRAS.492.4409M} {492, 4409}

\bibitem[\protect\citeauthoryear{{Marois}, {Macintosh}, {Barman}, {Zuckerman},
  {Song}, {Patience}, {Lafreni{\`e}re}  \& {Doyon}}{{Marois}
  et~al.}{2008}]{Marois2008}
{Marois} C.,  {Macintosh} B.,  {Barman} T.,  {Zuckerman} B.,  {Song} I.,
  {Patience} J.,  {Lafreni{\`e}re} D.,   {Doyon} R.,  2008, \mn@doi [Science]
  {10.1126/science.1166585}, \href
  {https://ui.adsabs.harvard.edu/abs/2008Sci...322.1348M} {322, 1348}

\bibitem[\protect\citeauthoryear{{Matr{\`a}}, {Wilner}, {{\"O}berg}, {Andrews},
  {Loomis}, {Wyatt}  \& {Dent}}{{Matr{\`a}} et~al.}{2018}]{Matra2018}
{Matr{\`a}} L.,  {Wilner} D.~J.,  {{\"O}berg} K.~I.,  {Andrews} S.~M.,
  {Loomis} R.~A.,  {Wyatt} M.~C.,   {Dent} W.~R.~F.,  2018, \mn@doi [\apj]
  {10.3847/1538-4357/aaa42a}, \href
  {https://ui.adsabs.harvard.edu/abs/2018ApJ...853..147M} {853, 147}

\bibitem[\protect\citeauthoryear{{Medina}, {Johnson}, {Eastman}  \&
  {Cargile}}{{Medina} et~al.}{2018}]{Medina2018}
{Medina} A.~A.,  {Johnson} J.~A.,  {Eastman} J.~D.,   {Cargile} P.~A.,  2018,
  \mn@doi [\apj] {10.3847/1538-4357/aadf82}, \href
  {https://ui.adsabs.harvard.edu/abs/2018ApJ...867...32M} {867, 32}

\bibitem[\protect\citeauthoryear{{Mendigut{\'\i}a}, {Calvet}, {Montesinos},
  {Mora}, {Muzerolle}, {Eiroa}, {Oudmaijer}  \& {Mer{\'\i}n}}{{Mendigut{\'\i}a}
  et~al.}{2011}]{Mendigutia2011b}
{Mendigut{\'\i}a} I.,  {Calvet} N.,  {Montesinos} B.,  {Mora} A.,  {Muzerolle}
  J.,  {Eiroa} C.,  {Oudmaijer} R.~D.,   {Mer{\'\i}n} B.,  2011, \mn@doi [\aap]
  {10.1051/0004-6361/201117444}, \href
  {https://ui.adsabs.harvard.edu/abs/2011A&A...535A..99M} {535, A99}

\bibitem[\protect\citeauthoryear{{Michel}, {van der Marel}  \&
  {Matthews}}{{Michel} et~al.}{2021}]{Michel2021}
{Michel} A.,  {van der Marel} N.,   {Matthews} B.~C.,  2021, \mn@doi [\apj]
  {10.3847/1538-4357/ac1bbb}, \href
  {https://ui.adsabs.harvard.edu/abs/2021ApJ...921...72M} {921, 72}

\bibitem[\protect\citeauthoryear{{Miley}, {Pani{\'c}}, {Wyatt}  \&
  {Kennedy}}{{Miley} et~al.}{2018}]{Miley2018}
{Miley} J.~M.,  {Pani{\'c}} O.,  {Wyatt} M.,   {Kennedy} G.~M.,  2018, \mn@doi
  [\aap] {10.1051/0004-6361/201833381}, \href
  {https://ui.adsabs.harvard.edu/abs/2018A&A...615L..10M} {615, L10}

\bibitem[\protect\citeauthoryear{{Mizusawa}, {Rebull}, {Stauffer}, {Bryden},
  {Meyer}  \& {Song}}{{Mizusawa} et~al.}{2012}]{Mizusawa2012}
{Mizusawa} T.~F.,  {Rebull} L.~M.,  {Stauffer} J.~R.,  {Bryden} G.,  {Meyer}
  M.,   {Song} I.,  2012, \mn@doi [\aj] {10.1088/0004-6256/144/5/135}, \href
  {https://ui.adsabs.harvard.edu/abs/2012AJ....144..135M} {144, 135}

\bibitem[\protect\citeauthoryear{{Modigliani} et~al.,}{{Modigliani}
  et~al.}{2010}]{Modigliani2010}
{Modigliani} A.,  et~al., 2010, in {Silva} D.~R.,  {Peck} A.~B.,   {Soifer}
  B.~T.,  eds,  Society of Photo-Optical Instrumentation Engineers (SPIE)
  Conference Series Vol. 7737, Observatory Operations: Strategies, Processes,
  and Systems III. p. 773728, \mn@doi{10.1117/12.857211}

\bibitem[\protect\citeauthoryear{{Moe} \& {Kratter}}{{Moe} \&
  {Kratter}}{2021}]{Moe2021}
{Moe} M.,  {Kratter} K.~M.,  2021, \mn@doi [\mnras] {10.1093/mnras/stab2328},
  \href {https://ui.adsabs.harvard.edu/abs/2021MNRAS.507.3593M} {507, 3593}

\bibitem[\protect\citeauthoryear{{Monet} et~al.,}{{Monet}
  et~al.}{2003}]{Monet2003}
{Monet} D.~G.,  et~al., 2003, \mn@doi [\aj] {10.1086/345888}, \href
  {https://ui.adsabs.harvard.edu/abs/2003AJ....125..984M} {125, 984}

\bibitem[\protect\citeauthoryear{{Mo{\'o}r} et~al.,}{{Mo{\'o}r}
  et~al.}{2017}]{Moor2017}
{Mo{\'o}r} A.,  et~al., 2017, \mn@doi [\apj] {10.3847/1538-4357/aa8e4e}, \href
  {https://ui.adsabs.harvard.edu/abs/2017ApJ...849..123M} {849, 123}

\bibitem[\protect\citeauthoryear{{Mulders}, {Pascucci}, {Apai}  \&
  {Ciesla}}{{Mulders} et~al.}{2018}]{Mulders2018}
{Mulders} G.~D.,  {Pascucci} I.,  {Apai} D.,   {Ciesla} F.~J.,  2018, \mn@doi
  [\aj] {10.3847/1538-3881/aac5ea}, \href
  {https://ui.adsabs.harvard.edu/abs/2018AJ....156...24M} {156, 24}

\bibitem[\protect\citeauthoryear{{Murphy} \& {Lawson}}{{Murphy} \&
  {Lawson}}{2015}]{Murphy_Lawson2015}
{Murphy} S.~J.,  {Lawson} W.~A.,  2015, \mn@doi [\mnras]
  {10.1093/mnras/stu2450}, \href
  {https://ui.adsabs.harvard.edu/abs/2015MNRAS.447.1267M} {447, 1267}

\bibitem[\protect\citeauthoryear{{Nakatani}, {Kobayashi}, {Kuiper}, {Nomura}
  \& {Aikawa}}{{Nakatani} et~al.}{2021}]{Nakatani2021}
{Nakatani} R.,  {Kobayashi} H.,  {Kuiper} R.,  {Nomura} H.,   {Aikawa} Y.,
  2021, \mn@doi [\apj] {10.3847/1538-4357/ac0137}, \href
  {https://ui.adsabs.harvard.edu/abs/2021ApJ...915...90N} {915, 90}

\bibitem[\protect\citeauthoryear{{Nelson} \& {Papaloizou}}{{Nelson} \&
  {Papaloizou}}{2004}]{Nelson2004}
{Nelson} R.~P.,  {Papaloizou} J. C.~B.,  2004, \mn@doi [\mnras]
  {10.1111/j.1365-2966.2004.07406.x}, \href
  {https://ui.adsabs.harvard.edu/abs/2004MNRAS.350..849N} {350, 849}

\bibitem[\protect\citeauthoryear{{Paardekooper} \& {Papaloizou}}{{Paardekooper}
  \& {Papaloizou}}{2009}]{Paardekooper2009}
{Paardekooper} S.~J.,  {Papaloizou} J.~C.~B.,  2009, \mn@doi [\mnras]
  {10.1111/j.1365-2966.2009.14511.x}, \href
  {https://ui.adsabs.harvard.edu/abs/2009MNRAS.394.2283P} {394, 2283}

\bibitem[\protect\citeauthoryear{{Paardekooper}, {Baruteau}, {Crida}  \&
  {Kley}}{{Paardekooper} et~al.}{2010}]{Paardekooper2010}
{Paardekooper} S.~J.,  {Baruteau} C.,  {Crida} A.,   {Kley} W.,  2010, \mn@doi
  [\mnras] {10.1111/j.1365-2966.2009.15782.x}, \href
  {https://ui.adsabs.harvard.edu/abs/2010MNRAS.401.1950P} {401, 1950}

\bibitem[\protect\citeauthoryear{{Palla} \& {Stahler}}{{Palla} \&
  {Stahler}}{1990}]{PallaStahler1990}
{Palla} F.,  {Stahler} S.~W.,  1990, \mn@doi [\apjl] {10.1086/185809}, \href
  {https://ui.adsabs.harvard.edu/abs/1990ApJ...360L..47P} {360, L47}

\bibitem[\protect\citeauthoryear{{Papaloizou}, {Nelson}, {Kley}, {Masset}  \&
  {Artymowicz}}{{Papaloizou} et~al.}{2007}]{Papaloizou2007}
{Papaloizou} J.~C.~B.,  {Nelson} R.~P.,  {Kley} W.,  {Masset} F.~S.,
  {Artymowicz} P.,  2007, in {Reipurth} B.,  {Jewitt} D.,   {Keil} K.,  eds,
  Protostars and Planets V. p.~655 (\mn@eprint {arXiv} {astro-ph/0603196})

\bibitem[\protect\citeauthoryear{{Pascucci} et~al.,}{{Pascucci}
  et~al.}{2016}]{Pascucci2016}
{Pascucci} I.,  et~al., 2016, \mn@doi [\apj] {10.3847/0004-637X/831/2/125},
  \href {https://ui.adsabs.harvard.edu/abs/2016ApJ...831..125P} {831, 125}

\bibitem[\protect\citeauthoryear{{Pearce} et~al.,}{{Pearce}
  et~al.}{2022}]{Pearce2022a}
{Pearce} T.~D.,  et~al., 2022, \mn@doi [\aap] {10.1051/0004-6361/202142720},
  \href {https://ui.adsabs.harvard.edu/abs/2022A&A...659A.135P} {659, A135}

\bibitem[\protect\citeauthoryear{{Pecaut} \& {Mamajek}}{{Pecaut} \&
  {Mamajek}}{2016}]{PecautMamajek2016}
{Pecaut} M.~J.,  {Mamajek} E.~E.,  2016, \mn@doi [\mnras]
  {10.1093/mnras/stw1300}, \href
  {https://ui.adsabs.harvard.edu/abs/2016MNRAS.461..794P} {461, 794}

\bibitem[\protect\citeauthoryear{{P{\'e}ricaud}, {Di Folco}, {Dutrey},
  {Guilloteau}  \& {Pi{\'e}tu}}{{P{\'e}ricaud} et~al.}{2017}]{Pericaud2017}
{P{\'e}ricaud} J.,  {Di Folco} E.,  {Dutrey} A.,  {Guilloteau} S.,
  {Pi{\'e}tu} V.,  2017, \mn@doi [\aap] {10.1051/0004-6361/201629371}, \href
  {https://ui.adsabs.harvard.edu/abs/2017A&A...600A..62P} {600, A62}

\bibitem[\protect\citeauthoryear{{Pinilla}, {Garufi}  \&
  {G{\'a}rate}}{{Pinilla} et~al.}{2022}]{Pinilla2022}
{Pinilla} P.,  {Garufi} A.,   {G{\'a}rate} M.,  2022, \mn@doi [\aap]
  {10.1051/0004-6361/202243637}, \href
  {https://ui.adsabs.harvard.edu/abs/2022A&A...662L...8P} {662, L8}

\bibitem[\protect\citeauthoryear{{Pollack}, {Hubickyj}, {Bodenheimer},
  {Lissauer}, {Podolak}  \& {Greenzweig}}{{Pollack} et~al.}{1996}]{Pollack1996}
{Pollack} J.~B.,  {Hubickyj} O.,  {Bodenheimer} P.,  {Lissauer} J.~J.,
  {Podolak} M.,   {Greenzweig} Y.,  1996, \mn@doi [\icarus]
  {10.1006/icar.1996.0190}, \href
  {https://ui.adsabs.harvard.edu/abs/1996Icar..124...62P} {124, 62}

\bibitem[\protect\citeauthoryear{{Prato}, {Greene}  \& {Simon}}{{Prato}
  et~al.}{2003}]{Prato2003}
{Prato} L.,  {Greene} T.~P.,   {Simon} M.,  2003, \mn@doi [\apj]
  {10.1086/345828}, \href
  {https://ui.adsabs.harvard.edu/abs/2003ApJ...584..853P} {584, 853}

\bibitem[\protect\citeauthoryear{{Reffert}, {Bergmann}, {Quirrenbach},
  {Trifonov}  \& {K{\"u}nstler}}{{Reffert} et~al.}{2015}]{Reffert2015}
{Reffert} S.,  {Bergmann} C.,  {Quirrenbach} A.,  {Trifonov} T.,
  {K{\"u}nstler} A.,  2015, \mn@doi [\aap] {10.1051/0004-6361/201322360}, \href
  {https://ui.adsabs.harvard.edu/abs/2015A&A...574A.116R} {574, A116}

\bibitem[\protect\citeauthoryear{{Ribas}, {Bouy}  \& {Mer{\'\i}n}}{{Ribas}
  et~al.}{2015}]{Ribas2015}
{Ribas} {\'A}.,  {Bouy} H.,   {Mer{\'\i}n} B.,  2015, \mn@doi [\aap]
  {10.1051/0004-6361/201424846}, \href
  {https://ui.adsabs.harvard.edu/abs/2015A&A...576A..52R} {576, A52}

\bibitem[\protect\citeauthoryear{{Rivinius}, {Carciofi}  \&
  {Martayan}}{{Rivinius} et~al.}{2013}]{Rivinius2013}
{Rivinius} T.,  {Carciofi} A.~C.,   {Martayan} C.,  2013, \mn@doi [\aapr]
  {10.1007/s00159-013-0069-0}, \href
  {https://ui.adsabs.harvard.edu/abs/2013A&ARv..21...69R} {21, 69}

\bibitem[\protect\citeauthoryear{{Royer}, {Gerbaldi}, {Faraggiana}  \&
  {G{\'o}mez}}{{Royer} et~al.}{2002}]{Royer2002}
{Royer} F.,  {Gerbaldi} M.,  {Faraggiana} R.,   {G{\'o}mez} A.~E.,  2002,
  \mn@doi [\aap] {10.1051/0004-6361:20011422}, \href
  {https://ui.adsabs.harvard.edu/abs/2002A&A...381..105R} {381, 105}

\bibitem[\protect\citeauthoryear{{Sbordone}, {Bonifacio}, {Castelli}  \&
  {Kurucz}}{{Sbordone} et~al.}{2004}]{Sbordone04}
{Sbordone} L.,  {Bonifacio} P.,  {Castelli} F.,   {Kurucz} R.~L.,  2004,
  Memorie della Societa Astronomica Italiana Supplementi, \href
  {http://adsabs.harvard.edu/abs/2004MSAIS...5...93S} {5, 93}

\bibitem[\protect\citeauthoryear{{Siess}, {Dufour}  \& {Forestini}}{{Siess}
  et~al.}{2000}]{Siess2000}
{Siess} L.,  {Dufour} E.,   {Forestini} M.,  2000, \aap, \href
  {http://adsabs.harvard.edu/abs/2000A%26A...358..593S} {358, 593}

\bibitem[\protect\citeauthoryear{{Skrutskie} et~al.,}{{Skrutskie}
  et~al.}{2006}]{Skrutskie2006}
{Skrutskie} M.~F.,  et~al., 2006, \mn@doi [\aj] {10.1086/498708}, \href
  {https://ui.adsabs.harvard.edu/abs/2006AJ....131.1163S} {131, 1163}

\bibitem[\protect\citeauthoryear{{Smette} et~al.,}{{Smette}
  et~al.}{2015}]{Smette2015}
{Smette} A.,  et~al., 2015, \mn@doi [\aap] {10.1051/0004-6361/201423932}, \href
  {http://adsabs.harvard.edu/abs/2015A%26A...576A..77S} {576, A77}

\bibitem[\protect\citeauthoryear{{Smirnov-Pinchukov}, {Mo{\'o}r}, {Semenov},
  {{\'A}brah{\'a}m}, {Henning}, {K{\'o}sp{\'a}l}, {Hughes}  \& {di
  Folco}}{{Smirnov-Pinchukov} et~al.}{2022}]{Smirnov-Pinchukov2022}
{Smirnov-Pinchukov} G.~V.,  {Mo{\'o}r} A.,  {Semenov} D.~A.,  {{\'A}brah{\'a}m}
  P.,  {Henning} T.,  {K{\'o}sp{\'a}l} {\'A}.,  {Hughes} A.~M.,   {di Folco}
  E.,  2022, \mn@doi [\mnras] {10.1093/mnras/stab3146}, \href
  {https://ui.adsabs.harvard.edu/abs/2022MNRAS.510.1148S} {510, 1148}

\bibitem[\protect\citeauthoryear{{Soderblom}, {Hillenbrand}, {Jeffries},
  {Mamajek}  \& {Naylor}}{{Soderblom} et~al.}{2014}]{Soderblom2014}
{Soderblom} D.~R.,  {Hillenbrand} L.~A.,  {Jeffries} R.~D.,  {Mamajek} E.~E.,
  {Naylor} T.,  2014, in {Beuther} H.,  {Klessen} R.~S.,  {Dullemond} C.~P.,
  {Henning} T.,  eds, Protostars and Planets VI. p.~219 (\mn@eprint {arXiv}
  {1311.7024}), \mn@doi{10.2458/azu\_uapress\_9780816531240-ch010}

\bibitem[\protect\citeauthoryear{{Stapper}, {Hogerheijde}, {van Dishoeck}  \&
  {Mentel}}{{Stapper} et~al.}{2022}]{Stapper2022}
{Stapper} L.~M.,  {Hogerheijde} M.~R.,  {van Dishoeck} E.~F.,   {Mentel} R.,
  2022, \mn@doi [\aap] {10.1051/0004-6361/202142164}, \href
  {https://ui.adsabs.harvard.edu/abs/2022A&A...658A.112S} {658, A112}

\bibitem[\protect\citeauthoryear{{The}, {de Winter}  \& {Perez}}{{The}
  et~al.}{1994}]{The1994}
{The} P.~S.,  {de Winter} D.,   {Perez} M.~R.,  1994, \aaps, \href
  {https://ui.adsabs.harvard.edu/abs/1994A&AS..104..315T} {104, 315}

\bibitem[\protect\citeauthoryear{{Torres}}{{Torres}}{2020}]{Torres2020}
{Torres} G.,  2020, \mn@doi [\apj] {10.3847/1538-4357/abb136}, \href
  {https://ui.adsabs.harvard.edu/abs/2020ApJ...901...91T} {901, 91}

\bibitem[\protect\citeauthoryear{\VAN{Leeuwen}{Van}{van}~Leeuwen}{\VAN{Leeuwen}{Van}{van}~Leeuwen}{2007}]{VanLeeuwen2007}
\VAN{Leeuwen}{Van}{van}~Leeuwen F.,  2007, \mn@doi [\aap]
  {10.1051/0004-6361:20078357}, \href
  {http://adsabs.harvard.edu/abs/2007A%26A...474..653V} {474, 653}

\bibitem[\protect\citeauthoryear{\VAN{den Ancker}{Van}{van}~den Ancker
  et~al.,}{\VAN{den Ancker}{Van}{van}~den Ancker
  et~al.}{2021}]{vandenAncker2021}
\VAN{den Ancker}{Van}{van}~den Ancker M.~E.,  et~al., 2021, \mn@doi [\aap]
  {10.1051/0004-6361/202141070}, \href
  {https://ui.adsabs.harvard.edu/abs/2021A&A...651L..11V} {651, L11}

\bibitem[\protect\citeauthoryear{{Vernet} et~al.,}{{Vernet}
  et~al.}{2011}]{Vernet2011}
{Vernet} J.,  et~al., 2011, \mn@doi [\aap] {10.1051/0004-6361/201117752}, \href
  {https://ui.adsabs.harvard.edu/abs/2011A&A...536A.105V} {536, A105}

\bibitem[\protect\citeauthoryear{{Vican}}{{Vican}}{2012}]{Vican2012}
{Vican} L.,  2012, \mn@doi [\aj] {10.1088/0004-6256/143/6/135}, \href
  {https://ui.adsabs.harvard.edu/abs/2012AJ....143..135V} {143, 135}

\bibitem[\protect\citeauthoryear{{Vieira}, {Corradi}, {Alencar}, {Mendes},
  {Torres}, {Quast}, {Guimar{\~a}es}  \& {da Silva}}{{Vieira}
  et~al.}{2003}]{Vieira2003}
{Vieira} S.~L.~A.,  {Corradi} W.~J.~B.,  {Alencar} S.~H.~P.,  {Mendes}
  L.~T.~S.,  {Torres} C.~A.~O.,  {Quast} G.~R.,  {Guimar{\~a}es} M.~M.,   {da
  Silva} L.,  2003, \mn@doi [\aj] {10.1086/379553}, \href
  {https://ui.adsabs.harvard.edu/abs/2003AJ....126.2971V} {126, 2971}

\bibitem[\protect\citeauthoryear{{Vioque}, {Oudmaijer}, {Baines},
  {Mendigut{\'\i}a}  \& {P{\'e}rez-Mart{\'\i}nez}}{{Vioque}
  et~al.}{2018}]{Vioque2018}
{Vioque} M.,  {Oudmaijer} R.~D.,  {Baines} D.,  {Mendigut{\'\i}a} I.,
  {P{\'e}rez-Mart{\'\i}nez} R.,  2018, \mn@doi [\aap]
  {10.1051/0004-6361/201832870}, \href
  {https://ui.adsabs.harvard.edu/abs/2018A&A...620A.128V} {620, A128}

\bibitem[\protect\citeauthoryear{{Vioque}, {Oudmaijer}, {Schreiner},
  {Mendigut{\'\i}a}, {Baines}, {Mowlavi}  \&
  {P{\'e}rez-Mart{\'\i}nez}}{{Vioque} et~al.}{2020}]{Vioque2020}
{Vioque} M.,  {Oudmaijer} R.~D.,  {Schreiner} M.,  {Mendigut{\'\i}a} I.,
  {Baines} D.,  {Mowlavi} N.,   {P{\'e}rez-Mart{\'\i}nez} R.,  2020, \mn@doi
  [\aap] {10.1051/0004-6361/202037731}, \href
  {https://ui.adsabs.harvard.edu/abs/2020A&A...638A..21V} {638, A21}

\bibitem[\protect\citeauthoryear{{Vioque} et~al.,}{{Vioque}
  et~al.}{2022}]{Vioque2022}
{Vioque} M.,  et~al., 2022, \mn@doi [\apj] {10.3847/1538-4357/ac5c46}, \href
  {https://ui.adsabs.harvard.edu/abs/2022ApJ...930...39V} {930, 39}

\bibitem[\protect\citeauthoryear{{Wagner}, {Apai}, {Kasper}, {McClure}  \&
  {Robberto}}{{Wagner} et~al.}{2022}]{Wagner2022}
{Wagner} K.,  {Apai} D.,  {Kasper} M.,  {McClure} M.,   {Robberto} M.,  2022,
  \mn@doi [\aj] {10.3847/1538-3881/ac409d}, \href
  {https://ui.adsabs.harvard.edu/abs/2022AJ....163...80W} {163, 80}

\bibitem[\protect\citeauthoryear{{Wenger} et~al.,}{{Wenger}
  et~al.}{2000}]{Wenger2000}
{Wenger} M.,  et~al., 2000, \mn@doi [\aaps] {10.1051/aas:2000332}, \href
  {http://adsabs.harvard.edu/abs/2000A%26AS..143....9W} {143, 9}

\bibitem[\protect\citeauthoryear{{Wichittanakom}, {Oudmaijer}, {Fairlamb},
  {Mendigut{\'\i}a}, {Vioque}  \& {Ababakr}}{{Wichittanakom}
  et~al.}{2020}]{Wichittanakom2020}
{Wichittanakom} C.,  {Oudmaijer} R.~D.,  {Fairlamb} J.~R.,  {Mendigut{\'\i}a}
  I.,  {Vioque} M.,   {Ababakr} K.~M.,  2020, \mn@doi [\mnras]
  {10.1093/mnras/staa169}, \href
  {https://ui.adsabs.harvard.edu/abs/2020MNRAS.493..234W} {493, 234}

\bibitem[\protect\citeauthoryear{{Wisniewski} \& {Bjorkman}}{{Wisniewski} \&
  {Bjorkman}}{2006}]{Wisniewski_Bjorkman2006}
{Wisniewski} J.~P.,  {Bjorkman} K.~S.,  2006, \mn@doi [\apj] {10.1086/507260},
  \href {https://ui.adsabs.harvard.edu/abs/2006ApJ...652..458W} {652, 458}

\bibitem[\protect\citeauthoryear{{Woosley}, {Heger}  \& {Weaver}}{{Woosley}
  et~al.}{2002}]{Woosley2002}
{Woosley} S.~E.,  {Heger} A.,   {Weaver} T.~A.,  2002, \mn@doi [Reviews of
  Modern Physics] {10.1103/RevModPhys.74.1015}, \href
  {https://ui.adsabs.harvard.edu/abs/2002RvMP...74.1015W} {74, 1015}

\bibitem[\protect\citeauthoryear{{Wright} \& {Drake}}{{Wright} \&
  {Drake}}{2016}]{WrightDrake2016}
{Wright} N.~J.,  {Drake} J.~J.,  2016, \mn@doi [\nat] {10.1038/nature18638},
  \href {https://ui.adsabs.harvard.edu/abs/2016Natur.535..526W} {535, 526}

\bibitem[\protect\citeauthoryear{{Wright}, {Egan}, {Kraemer}  \&
  {Price}}{{Wright} et~al.}{2003}]{Wright2003}
{Wright} C.~O.,  {Egan} M.~P.,  {Kraemer} K.~E.,   {Price} S.~D.,  2003,
  \mn@doi [\aj] {10.1086/345511}, \href
  {https://ui.adsabs.harvard.edu/abs/2003AJ....125..359W} {125, 359}

\bibitem[\protect\citeauthoryear{{Wright} et~al.,}{{Wright}
  et~al.}{2010}]{Wright2010}
{Wright} E.~L.,  et~al., 2010, \mn@doi [\aj] {10.1088/0004-6256/140/6/1868},
  \href {https://ui.adsabs.harvard.edu/abs/2010AJ....140.1868W} {140, 1868}

\bibitem[\protect\citeauthoryear{{Wyatt}}{{Wyatt}}{2008}]{Wyatt2008}
{Wyatt} M.~C.,  2008, \mn@doi [\araa] {10.1146/annurev.astro.45.051806.110525},
  \href {http://adsabs.harvard.edu/abs/2008ARA%26A..46..339W} {46, 339}

\bibitem[\protect\citeauthoryear{{Wyatt}, {Pani{\'c}}, {Kennedy}  \&
  {Matr{\`a}}}{{Wyatt} et~al.}{2015}]{Wyatt2015}
{Wyatt} M.~C.,  {Pani{\'c}} O.,  {Kennedy} G.~M.,   {Matr{\`a}} L.,  2015,
  \mn@doi [\apss] {10.1007/s10509-015-2315-6}, \href
  {http://adsabs.harvard.edu/abs/2015Ap%26SS.357..103W} {357, 103}

\bibitem[\protect\citeauthoryear{{Yamada} \& {Inaba}}{{Yamada} \&
  {Inaba}}{2011}]{YamadaInaba2011}
{Yamada} K.,  {Inaba} S.,  2011, \mn@doi [\mnras]
  {10.1111/j.1365-2966.2010.17670.x}, \href
  {https://ui.adsabs.harvard.edu/abs/2011MNRAS.411..184Y} {411, 184}

\bibitem[\protect\citeauthoryear{{Yang}, {Xie}  \& {Zhou}}{{Yang}
  et~al.}{2020}]{Yang2020}
{Yang} J.-Y.,  {Xie} J.-W.,   {Zhou} J.-L.,  2020, \mn@doi [\aj]
  {10.3847/1538-3881/ab7373}, \href
  {https://ui.adsabs.harvard.edu/abs/2020AJ....159..164Y} {159, 164}

\bibitem[\protect\citeauthoryear{{York} et~al.,}{{York}
  et~al.}{2000}]{York2000}
{York} D.~G.,  et~al., 2000, \mn@doi [\aj] {10.1086/301513}, \href
  {https://ui.adsabs.harvard.edu/abs/2000AJ....120.1579Y} {120, 1579}

\bibitem[\protect\citeauthoryear{{Zorec}, {Fr{\'e}mat}  \& {Cidale}}{{Zorec}
  et~al.}{2005}]{Zorec2005}
{Zorec} J.,  {Fr{\'e}mat} Y.,   {Cidale} L.,  2005, \mn@doi [\aap]
  {10.1051/0004-6361:20053051}, \href
  {https://ui.adsabs.harvard.edu/abs/2005A&A...441..235Z} {441, 235}

\makeatother
\end{thebibliography}




\appendix

\section{Herbig Ae/Be Stars in the sample and Accretion diagnostics}
\label{sec:accretion}

Herbig Ae/Be stars are pre-main-sequence intermediate mass stars characterized by the presence of emission lines and an IR excess \citep{Hillenbrand1992, Hernandez2004}. Since our initial sample consists of pre-main-sequence IMS candidates, it is likely that some of them are Herbig Ae/Be stars. Therefore, we studied accretion tracers in the full sample looking for emission lines indicative of falling into the category of Herbig Ae/Be stars. We found a total of six stars among these candidates presenting emission in lines such as the Balmer series, Br$\gamma$, Pa$\alpha$ and Pa$\beta$ among others. These six Herbig stars were previously identified in the literature (e.g. in \citealt{The1994}, \citealt{Vieira2003}, \citealt{Chen2016}, \citealt{Vioque2018,Vioque2022}, \citealt{Arun2019}, \citealt{Wichittanakom2020}, \citealt{GuzmanDiaz2021}), however, we wanted to characterize them as well as part of our sample and present them in comparison to other studies. 

\begin{figure}
	\includegraphics[width=\columnwidth]{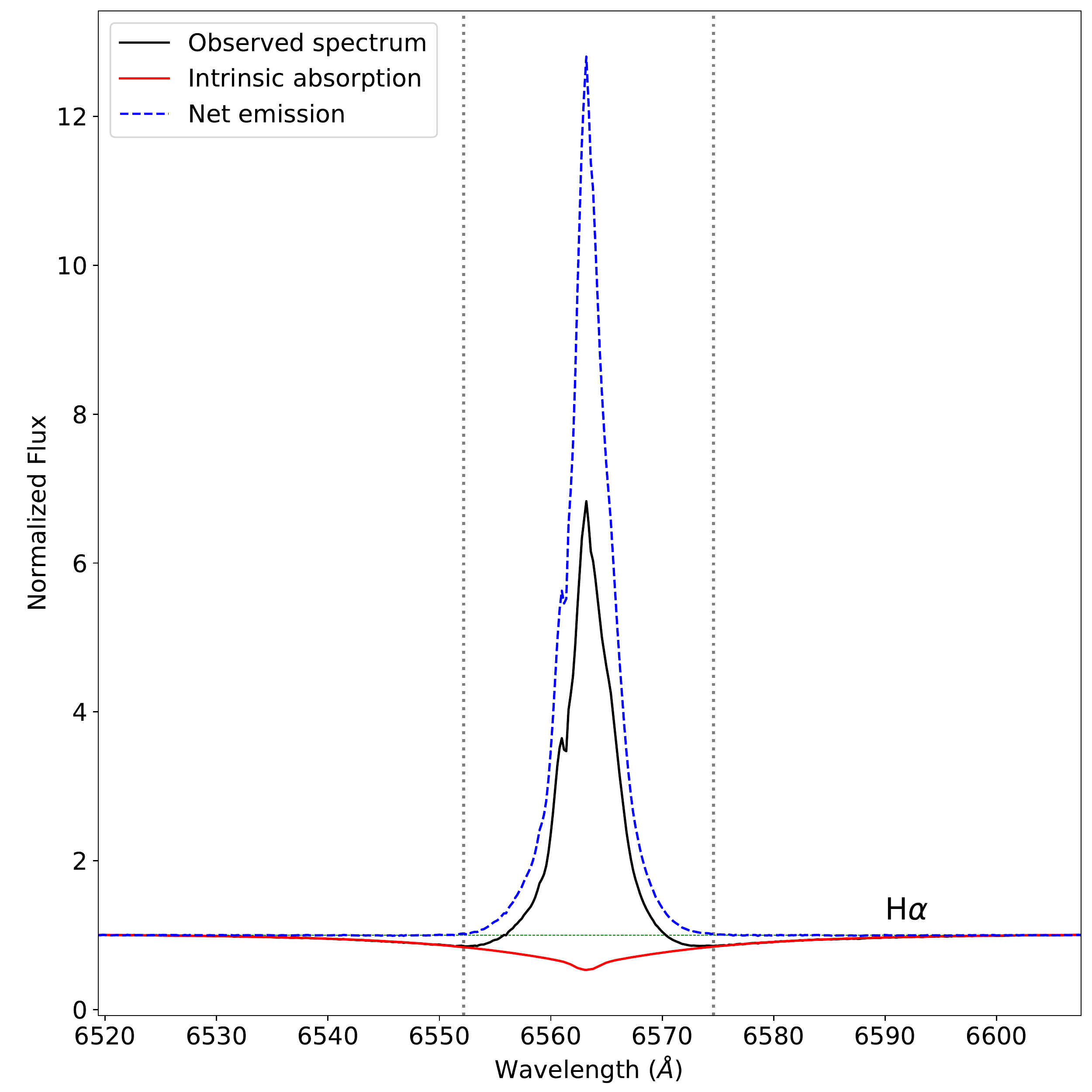}
    \caption{The observed H$\alpha$ line of HIP\,56379 (in black), the underlying absorption estimated by fitting a Kurucz model to the wings of the line (in red), and the net emission (in dashed blue line). The vertical dotted gray lines mark the limits of the central region that was excluded for the model fitting. The dashed green horizontal line marks the continuum level. The stellar parameters used to fit the underlying absorption of this particular line are: T$_{\rm eff}$=12500 K, $\log (g)$=4.7 dex and $v$sin$i$= 70 km s$^{-1}$. The resulting measurements of EW in this case were: EW$_{\rm obs}$=-23.24$\pm$0.92, EW$_{\rm abs}$=9.07$\pm$0.91, and EW$_{\rm emi}$=-32.32$\pm$1.29.}
    \label{fig:Halpha}
\end{figure}

Accretion luminosities and mass accretion rates can be estimated based on the intensity of the H$\alpha$ line, since this is the strongest gas accretion tracer (e.g. \citealt{Mendigutia2011b}). The strength of this emission line can be quantified by measuring its equivalent width. However, the observed equivalent width EW$_{\rm obs}$ is affected by the intrinsic absorption, the emission, and the contribution of any continuum excess at a given wavelength. For the case of the H$\alpha$ line the excess contribution is negligible, so the EW of the net emission can be estimated as EW$_{\rm emi}$ = EW$_{\rm obs}$ - EW$_{\rm abs}$, where EW$_{\rm abs}$ is the EW of the intrinsic absorption. To estimate the EW$_{\rm abs}$ we have used the Kurucz model that best fits the wings of the H$\alpha$ line excluding the central region containing the emission, as shown in the example in Figure \ref{fig:Halpha}. Since the error in the EW$_{\rm obs}$ is dominated by the noise of the target spectrum, the uncertainties for the EW$_{\rm obs}$ are estimated as the extent of the wavelength region used for the measurement divided by the SNR of the continuum: $\sigma_{\rm EW_{obs}}=\Delta \lambda$/SNR. The EW$_{\rm abs}$ depends mainly on the model used to estimate the intrinsic absorption and the shape of this model is based on a combination of stellar parameters. Given that different combinations of parameters will result on different model shapes, it is complicated to quantify the uncertainty exactly. Hence, since uncertainties for the stellar parameters are in the order of $\sim$10$\%$, we will adopt a (likely generous) 10$\%$ of uncertainty of the EW$_{\rm abs}$. The uncertainty of EW$_{\rm emi}$ is then given by error propagation. Although it is also possible to measure EW$_{\rm emi}$ directly by integrating the emission line profile of the net emission after subtracting the underlying absorption, by obtaining EW$_{\rm obs}$ and EW$_{\rm abs}$ independently we avoid introducing the uncertainty in radial velocity shift between the model and the observed spectrum, as they must be aligned for the subtraction.
\\

\begin{table*}
	\centering
	\caption{Measurements of equivalent widths, fluxes and mass accretion rates for the sub-sample of Herbig Ae/Be stars based on their H$\alpha$ emission.}
	\label{tab:Herbig}
	\begin{tabular*}{\textwidth}{lcccccccc} 
		\hline   
			Name & EW$_{\rm obs}$ & EW$_{\rm abs}$  &  EW$_{\rm emi}$  &  F$_{\lambda}$ & F$_{\rm emi}$  & $\log$(L$_{\rm emi}$) & $\log$(L$_{\rm acc}$) & $\log$($\dot M_{acc}$)      \\
			       & (\AA)  & (\AA) & (\AA)  &  (W m$^{-2}$\AA$^{-1}$) & (W m$^{-2}$) & [L$_{\odot}$] & [L$_{\odot}$] &  [M$_{\odot}$yr$^{-1}$]   \\
		\hline
HIP56379 & -23.24$\pm$0.92 & 9.07$\pm$0.91 & -32.32$\pm$1.29 & 6.43$\times 10^{-15}$ & (2.08$\pm$0.08)$\times 10^{-13}$ & -1.12$\pm^{0.02}_{0.02}$ & 0.97$\pm^{0.10}_{0.10}$ & -6.69$\pm^{0.03}_{0.03}$ \\
HIP78092 & -6.37$\pm$0.69 & 5.05$\pm$0.50 & -11.41$\pm$0.86 & 1.55$\times 10^{-15}$ & (1.77$\pm$0.13)$\times 10^{-14}$ & -1.86$\pm^{0.03}_{0.03}$ & 0.23$\pm^{0.12}_{0.12}$ & -7.09$\pm^{0.09}_{0.06}$ \\
HIP85755 & 3.90$\pm$0.34 & 7.38$\pm$0.74 & -3.47$\pm$0.81 & 3.81$\times 10^{-14}$ & (1.33$\pm$0.31)$\times 10^{-13}$ & -1.18$\pm^{0.09}_{0.12}$ & 0.91$\pm^{0.01}_{0.02}$ & -6.49$\pm^{0.08}_{0.07}$ \\
HIP87819 & -12.54$\pm$0.76 & 9.73$\pm$0.97 & -22.27$\pm$1.23 & 6.83$\times 10^{-15}$ & (1.52$\pm$0.08)$\times 10^{-13}$ & -1.32$\pm^{0.02}_{0.03}$ & 0.77$\pm^{0.10}_{0.10}$ & -6.82$\pm^{0.03}_{0.04}$ \\
HIP94260 & 1.19$\pm$0.90 & 10.00$\pm$1.00 & -8.81$\pm$1.35 & 4.00$\times 10^{-15}$ & (3.52$\pm$0.54)$\times 10^{-14}$ & -1.13$\pm^{0.06}_{0.07}$ & 0.96$\pm^{0.05}_{0.05}$ & -6.54$\pm^{0.05}_{0.03}$ \\
TYC6779-305-1 & -6.29$\pm$0.63 & 4.97$\pm$0.50 & -11.26$\pm$0.80 & 3.92$\times 10^{-16}$ & (4.42$\pm$0.32)$\times 10^{-15}$ & -2.42$\pm^{0.03}_{0.03}$ & -0.33$\pm^{0.15}_{0.15}$ & -7.77$\pm^{0.05}_{0.04}$ \\
		\hline
	\end{tabular*}
\end{table*}

The flux of the emission line is estimated as F$_{\rm emi}$ = EW$_{\rm emi}\times$F$_{\lambda}$, where F$_{\lambda}$ is the estimated continuum flux at the central wavelength of the line, in this case, H$\alpha$. F$_{\lambda}$ is calculated in a similar manner as other fluxes in previous sections; by using the Kurucz model with the stellar parameters and scaling it by the $D/R_{\star}$ factor. The emission line luminosity is calculated as L$_{\rm emi}$ = 4$\pi$D$^{2}$F$_{\rm emi}$, where D is the distance to the star. Uncertainties for F$_{\rm emi}$ and L$_{\rm emi}$ are derived by error propagation of the values involved.

The accretion luminosity is calculated in terms of the emission line luminosity by following the relationship \citep{Mendigutia2011b}:

\begin{equation}
    \log \left(\frac{L_{\rm acc}}{L_{\odot}}\right) = A + B\times\log \left(\frac{L_{\rm emi}}{L_{\odot}}\right)
\end{equation}

where $A$ and $B$ are constants that vary depending on the line. For the case of H$\alpha$, we use the relationship determined by \cite{Fairlamb2017}, where $A$ = 2.09 $\pm$ 0.06 and $B$ = 1.00 $\pm$ 0.05. Then, the mass accretion rate is determined from the accretion luminosity, stellar radius and stellar mass as \citep{Wichittanakom2020}:

\begin{equation}
    \dot M_{acc} = \frac{L_{\rm acc}R_{\star}}{G M_{\star}}
\end{equation}

\begin{figure}
	\includegraphics[width=\columnwidth]{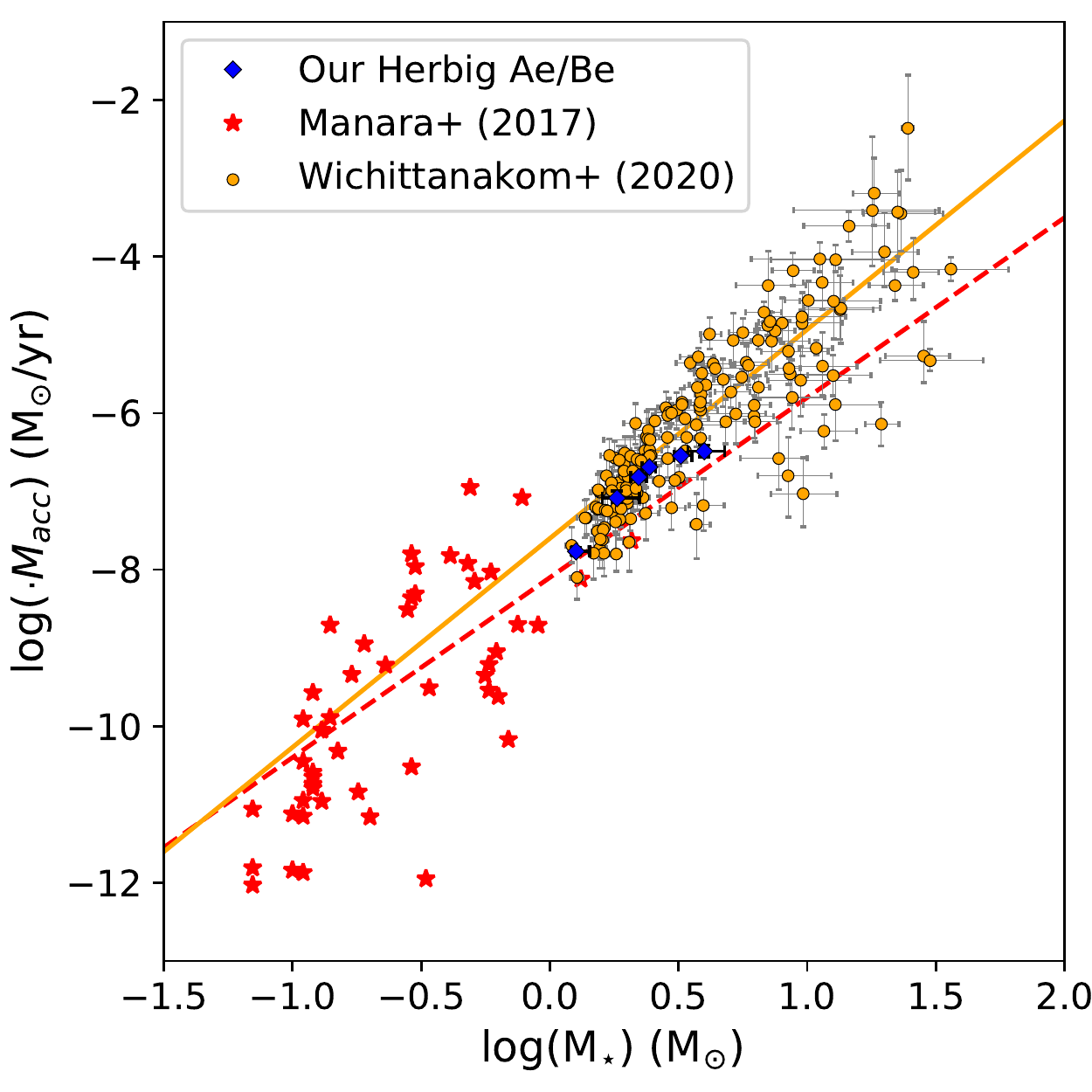}
    \caption{Mass accretion rates in our sample compared to other studies. Single-line fits are shown for reference; dashed red line for M17, and solid orange line for W20. Error bars might be smaller than markers.}
    \label{fig:Macc}
\end{figure}

Uncertainties for L$_{\rm acc}$ and $\dot M_{acc}$ are derived by error propagation as well. EW$_{\rm obs}$, EW$_{\rm abs}$, EW$_{\rm emi}$, F$_{\lambda}$, F$_{\rm emi}$, L$_{\rm emi}$, L$_{\rm acc}$ and $\dot M_{acc}$ for the 6 Herbig Ae/Be stars in the sample are presented in Table \ref{tab:Herbig}. The relationship between $M_{\star}$ and $\dot M_{acc}$ is shown in Figure \ref{fig:Macc}.

\subsection{Mass accretion rates}

Figure \ref{fig:Macc} shows the mass accretion rates $\dot M_{acc}$ for the  Herbig Ae/Be stars of our sample, along with the sample of LMSs in the Chamaeleon\,I star-forming region studied in \cite{Manara2017} (M17, their Fig. 4), and the full sample of Herbig Ae/Be stars studied in \cite{Wichittanakom2020} (W20, their Fig. 7, right). The best single-line fits from these works are shown, considering the full mass range in each case. Our sample of accreting stars consists only of six objects, which are also shown for comparison. Their masses and accretion rates are consistent with the trends in the literature, especially if we consider the limited number of objects and the fact that these studies also show different slopes when splitting their samples into lower and higher masses. Other works, such as \cite{Mendigutia2011b} find much steeper slopes for a range of masses similar to our sample.  

\section{Comments on individual objects}
\label{sec:comments}

\subsection{Peculiar objects}

\noindent

\noindent
\textbf{HIP\,82714 (HD\,152384):} This object was studied in detail by \cite{vandenAncker2021}. They found the presence of refractory elements such as Ca\,{\sc ii}, Fe\,{\sc i}, Mg\,{\sc i} and Si\,{\sc i}, and an absence of volatiles. They suggest this circumstellar material might be attributable to collisions of rocky planets. Interestingly, this target falls within our selection of young intermediate mass stars and within the `hybrid' disc classification.

\subsection{Contaminants}

\subsubsection{Classical Be stars}

As discussed in Section \ref{sec:Ages}, we found 16 emission line stars in the initial sample likely to be classical Be stars. For future reference, these stars are listed in Table \ref{tab:CBe}. Those falling within the IMSs refined selection (and excluded from it) are marked with a *. Those objects without previous classification as CBe stars in the literature are highlighted in bold face.

\begin{table}
	\centering
	\caption{List of contaminants: classical Be stars (CBe), giant stars, stars belonging to older associations, and stars with nebulosity contamination (in square brackets).}
	\label{tab:CBe}
	\begin{tabular}{ccc} 
		\hline
		CBe & Giants & In Assoc. \\
		\hline
		*HIP\,108402 & *HIP\,104305 & [HIP\,17527] \\
		HIP\,22112 & \textbf{*HIP\,23528} & [HIP\,17588] \\
		\textbf{HIP\,23201} & \textbf{HIP\,74242} & [HIP\,17664] \\
		HIP\,28561 & *HIP\,78378 & [HIP\,17862] \\
		HIP\,29635 & HIP\,79945 & [HIP\,19720] \\
		HIP\,30448 & \textbf{HIP\,85488} & HIP\,26966 \\
		HIP\,38779 & \textbf{*HIP\,86866} & HIP\,33359 \\
		HIP\,51491 & \textbf{HIP\,87422} & HIP\,52160 \\
		HIP\,57143 & \textbf{*TYC\,1235-186-1} & [HIP\,80019] \\
		HIP\,61738 & TYC\,1605-722-1 \\
		HIP\,74911 & \textbf{TYC\,462-1284-1} \\
		\textbf{HIP\,77289} & \textbf{*TYC\,463-4076-1} \\
		HIP\,81710 & \textbf{*TYC\,5116-72-1} \\
		\textbf{*HIP\,92364} & \textbf{TYC\,6858-2632-1} \\
		HIP\,94770 & \textbf{TYC\,7903-1637} \\
		*HIP\,95109 & \textbf{TYC\,8697-1523-1} \\
		& \textbf{TYC\,9394-1843-1} \\
		\hline
	\end{tabular}
\end{table}

\subsubsection{Giant stars}

Following on the age assessment in Section \ref{sec:Ages}, we also found a number of giant stars in the sample, seven of them within the refined sample were discarded. For reference, these stars are listed in Table \ref{tab:CBe}. Same as for the CBe stars, those objects without previous classification as giant stars in the literature are highlighted in bold face, and those falling in the refined selection are marked with a *.

\subsubsection{Stars belonging to older associations and stars with contamination}

As described in Section \ref{sec:Ages}, we found eight stars with $\geq$99\% probability of belonging to a stellar association with an estimated age in the literature that is incompatible with a pre-MS age. In addition, we found six stars presenting contamination from reflection nebulae, five of them among the ones found in associations older than 30Myrs. These stars are highlighted in square brackets in Table \ref{tab:CBe}. HIP\,17527, HIP\,17588, HIP\,17664, and HIP\,17862 belong to the Pleiades cluster, aged 112$\pm$5 Myrs \citep{Dahm2015}. These stars are located in the Merope Nebula, which is a diffuse reflection nebula, a cloud rich in interstellar dust reflecting the light of bright stars, and therefore these stars suffer from high extinction. Similarly, HIP\,19720 is found in the $\mu$ Tau association, a 62$\pm$7 Myr coeval group of stars \citep{Gagne2020} and it is also affected from contamination of a reflection nebula. HIP\,26966, HIP\,33359, and HIP\,52160 are found in the Columba association (42$\pm^{6}_{4}$ Myrs; \citealt{Bell2015}), Octans association (35$\pm$5 Myrs; \citealt{Murphy_Lawson2015}), and IC2602 open cluster (46$\pm^{6}_{5}$ Myrs; \citealt{Dobbie2010}), respectively. Finally, HIP\,80019 is found in the Upper Scorpius association (10$\pm$3 Myrs, \citealt{PecautMamajek2016}) and, although the estimated age for this star (7$\pm^{1}_{2}$ Myrs) is consistent with that of the stellar association, it presents contamination from a reflection nebula and thus discarded from the analysis as well.



\bsp	
\label{lastpage}
\end{document}